\documentclass[manuscript]{acmart}
\usepackage{algorithm}
\usepackage{algpseudocode}
\usepackage{tabularray}
\usepackage{bm}
\usepackage{hhline}
\usepackage{tabularx}
\usepackage{float}
\usepackage{placeins} 
\usepackage{graphicx}              
\usepackage{subcaption}            
\usepackage{color}
\usepackage[normalem]{ulem}
\AtBeginDocument{%
  }

\setcopyright{acmlicensed}
\copyrightyear{2018}
\acmYear{2018}
\acmDOI{XXXXXXX.XXXXXXX}
\acmConference[Conference acronym 'XX]{Make sure to enter the correct
  conference title from your rights confirmation email}{June 03--05,
  2018}{Woodstock, NY}
\acmISBN{978-1-4503-XXXX-X/2018/06}

\begin{document}

\title{Performance-Driven QUBO for Recommender Systems on Quantum Annealers}


\author{Jiayang Niu}
\email{s4068570@student.rmit.edu.au}
\affiliation{%
  \institution{School of Computing Technologies, RMIT University}
  \city{Melbourne}
  \country{Australia}
}

\author{Jie Li}
\email{hey.jieli@gmail.com}
\affiliation{%
  \institution{School of Computing Technologies, RMIT University}
  \city{Melbourne}
  \country{Australia}
}

\author{Ke Deng}
\email{ke.deng@rmit.edu.au}
\affiliation{%
  \institution{School of Computing Technologies, RMIT University}
  \city{Melbourne}
  \country{Australia}
}

\author{Mark Sanderson}
\email{mark.sanderson@rmit.edu.au}
\affiliation{%
  \institution{School of Computing Technologies, RMIT University}
  \city{Melbourne}
  \country{Australia}
}

\author{Nicola Ferro}
\email{nicola.ferro@unipd.it}
\affiliation{%
  \institution{Department of Information Engineering, University of Padua}
  \city{Padua}
  \country{Italy}
}

\author{Yongli Ren}
\email{yongli.ren@rmit.edu.au}
\affiliation{%
  \institution{School of Computing Technologies, RMIT University}
  \city{Melbourne}
  \country{Australia}
}

\renewcommand{\shortauthors}{Jiayang Niu et al.}

\begin{abstract}

Quantum annealers offer a promising hardware platform for solving combinatorial optimization problems, especially those formulated as Quadratic Unconstrained Binary Optimization (QUBO). In this work, we propose \textbf{PDQUBO} (\textbf{P}erformance-\textbf{D}riven \textbf{Q}uadratic \textbf{U}nconstrained \textbf{B}inary \textbf{O}ptimization), a QUBO-based feature selection method that is directly executable on quantum annealers. Unlike prior QUBO-based feature selection approaches on quantum annealers, PDQUBO explicitly quantifies the performance impact of both individual features and feature pairs on recommender system models. This alignment between QUBO optimization objectives and model performance ensures that the solution direction is closely tied to recommendation quality, making it well-suited for practical deployment on quantum hardware. Moreover, by leveraging counterfactual analysis, PDQUBO is model-agnostic and evaluation-metric-independent, making it broadly applicable across diverse recommender architectures and assessment criteria. In addition, we investigate the instability of quantum annealing on real quantum devices with respect to varying problem sizes and problem difficulties. Extensive experiments on real-world datasets demonstrate that PDQUBO consistently outperforms prior QUBO-based feature selection methods on quantum annealers. Furthermore, we compare PDQUBO against classical feature selection baselines on click-through rate (CTR) prediction tasks, showing its strong performance and highlighting the potential of using quantum annealers for real-world feature selection applications. Our findings suggest that integrating quantum optimization with counterfactual analysis provides a promising direction for effective feature selection in recommender systems.
\end{abstract}




\maketitle

\section{Introduction}

Compared to traditional transistor-based computers, quantum computers~\cite{wang2017experimental,chi2022programmable,ladd2010quantum} are still in the experimental stage and currently operate in the so-called Noisy Intermediate-Scale Quantum (NISQ) era~\cite{bharti2022noisy, niu2025estimating}. Despite hardware limitations such as noise and limited qubit counts, NISQ devices have shown promising capabilities in solving combinatorial optimisation problems, particularly due to their increased potential to escape local minima~\cite{PhysRevE.58.5355,yarkoni2022quantum,albash2018demonstration,DBLP:journals/corr/BelloniLNR15}. Recent studies show that quantum annealing (QA) can effectively escape local optima via quantum tunneling and explore vast solution spaces, making it well-suited for high-dimensional combinatorial optimisation~\cite{PhysRevE.58.5355,yarkoni2022quantum, albash2018demonstration,DBLP:journals/corr/BelloniLNR15}. This potential advantage has been theoretically supported by Tadashi and Hidetoshi~\cite{kadowaki1998quantum}, who demonstrated the superiority of QA over classical approaches under certain conditions.
Moreover, Quantum Annealing naturally aligns with Quadratic Unconstrained Binary Optimization (QUBO) problems~\cite{glover2018tutorial}, enabling us to leverage existing quantum annealers to explore combinatorial optimization problems that are challenging for classical computers to solve~\cite{pastorello2019quantum}. In parallel, feature selection has emerged as a critical challenge in modern recommender systems. 
Although embedding techniques~\cite{he2017neural, guo2017deepfm, he2020lightgcn} have mitigated certain feature engineering challenges in recommender systems, feature selection remains a critical issue. Selecting a compact and informative subset of features not only reduces storage requirements but also improves retrieval efficiency and system scalability. Recent studies~\cite{wang2022autofield, zhang2023shark, jia2024erase, guo2022lpfs, lin2022adafs, fisher2019all} have highlighted that redundant or less informative features can negatively affect both efficiency and performance. Moreover, prior work~\cite{faggioli2024dimension} has demonstrated that embedding representations often contain substantial redundancy, and that optimization-based methods such as QUBO can effectively identify and remove redundant dimensions in retrieval systems. These findings further underscore the importance of principled feature selection strategies in modern recommender systems.


One potential approach to overcoming these limitations is through QUBO-based feature selection. The Quadratic Unconstrained Binary Optimization (QUBO) problem aims to minimize a quadratic objective of the form $\min_{x \in \{0,1\}^n} x^\top Q x$, where $x$ is a binary vector and $Q$ is a coefficient matrix encoding individual contributions and pairwise interactions among variables. Such problems can be solved using quantum annealers, has been demonstrated in previous studies~\cite{nembrini2021feature, ferrari2022towards} to be highly suitable for feature selection problems, as it considers pairwise coupling relationships between features and identifies optimal feature combinations. Furthermore, due to the powerful advantages of quantum annealers in solving combinatorial problems, QUBO is regarded as one of the promising approaches to exploring feature selection problems~\cite{mucke2023feature}.However, there are significant challenges when using quantum annealing to solve complicated real world feature selection for recommendation purposes, and this paper investigates this by answering the following questions:

\begin{itemize}
    \item Q1) While having the potential, how does the instability of a quantum annealer affect the performance of recommender systems?
    \item Q2) How effective is a quantum annealer in feature selection for recommender systems?
    \item Q3) Can emerging QUBO-based feature selection methods on quantum annealers achieve, or even surpass, the performance of mature feature selection approaches on classical computers?
\end{itemize}

In this paper, following existing research, we investigate these questions in the context of QUBO-based feature selection for recommender systems on quantum annealers. 
Specifically, the QUBO formulation tailored to the feature selection problem in recommender systems is far from straightforward, and the key challenge lies in how to model the relationship between features and the optimization of recommendation problem~\cite{glover2018tutorial}. 
One approach is the construction of a Coefficient Matrix $Q$, whose detailed formulation is provided in Sec.~\ref{sec:QUBO}, which determines the alignment of the optimization direction of the QUBO problem with the optimization needs of feature selection. 

Existing research explored the construction of this coefficient matrix with the similarities between collaborative filtering and content-based models~\cite{nembrini2021feature} , 
mutual information, correlation, or the similarity between the classification predictions~\cite{ferrari2022towards}. 
However, although these existing works used either the ground truth (e.g. classification labels) or the model outcome (e.g. clarification predictions), none of them consider the performance of the models (e.g. accuracy in classification or recommender systems). 

So, to answer the above research questions, 
we propose a \textbf{PDQUBO} (\textbf{P}erformance-\textbf{D}riven \textbf{Q}uadratic \textbf{U}nconstrained \textbf{B}inary \textbf{O}ptimization).
The optimizer constructs the coefficient matrix by modeling the relationship between features and their impact on model performance via counterfactual analysis, which evaluates the changes in the final recommendation performance when excluding each feature and each pair of features, known as counterfactual instances. With counterfactual analysis, PDQUBO has two desirable properties. First, PDQUBO explicitly connects optimizing feature selection with the final recommendation performance rather than being based on either the ground truth or the model outcome only. This makes feature selection performance-driven, which contributes to its effectiveness in terms of generating good recommendations. Second, when modeling the importance of features in PDQUBO, the counterfactual analysis ensures the measure is their relative importance to all other features on recommendation performance, which can contribute to the performance in feature selection itself. Due to these properties, PDQUBO drives the optimization direction of the QUBO problem towards the optimization of recommendation performance. Note, with counterfactual analysis, PDQUBO is independent of what the recommender systems are used (known as \textit{base models}) and what recommendation metrics are employed to measure performance. Finally, given PDQUBO is performance-driven, it provides a unique chance to study the instability of quantum anneals and its performance.

The contributions of this paper are as follows:
\begin{itemize}
    \item The PDQUBO model, which aligns optimization direction to the recommendation performance and considers the hidden relationship among features to allow better optimization. 
    \item Comprehensive analysis of the instability of Quantum Annealers in feature selection for recommender systems, from the perspective of the number of features, the sample size in quantum annealing process, and the difficulties of the feature selection problems. 
    \item Extensive experiments on real-world datasets using various base models demonstrate that the proposed PDQUBO method significantly outperforms existing quantum annealer-based feature selection approaches, and achieves competitive or improved performance compared to traditional methods in most cases. 
\end{itemize}

The structure of this paper is as follows: Sec.\ref{sec:RW} reviews related work, with a focus on feature selection using quantum annealers and counterfactual analysis. Sec.\ref{sec:PRL} introduces the necessary preliminaries, including notation and the formal definition of QUBO problems. Sec.\ref{sec:METHOD} describes the proposed PDQUBO method in detail. Sec.\ref{sec:EXP} presents comprehensive experiments, including two main evaluation parts and an analysis of the stability of quantum annealers. Finally, Sec.\ref{sec:conclusion} concludes the paper and outlines potential future directions.

\section{RELATED WORK}
\label{sec:RW}
\subsection{Quantum Annealing}
\subsubsection{Recent applications:}
With the rapid advancement of quantum computing architectures~\cite{lloyd2014quantum, rebentrost2014quantum}, researchers have started to explore the potential of leveraging quantum computers’ computational advantages to address certain general problems, particularly in domains frequently confronted with NP-hard optimization challenges~\cite{boixo2016computational, zhu2022adaptive, bhatia2020quantum} and the increasing computational demands of deep learning~\cite{cong2019quantum, chalumuri2021hybrid, ajagekar2023molecular, date2021qubo}. Quantum annealers~\cite{wang2017experimental}, a type of quantum computer, are particularly suited for solving combinatorial optimization problems~\cite{chi2022programmable} and QUBO (Quadratic Unconstrained Binary Optimization) problems~\cite{kadowaki1998quantum}.Recent studies have focused on formalizing problems in these areas as QUBO problems to leverage quantum annealers for their solutions. \citet{date2021qubo} attempted to formalize simple machine learning methods as QUBO problems with quantum annealers. Some studies~\cite{alavi2025quantum, mucke2023feature, nembrini2021feature, ferrari2022towards} have also focused on utilizing quantum annealers to solve feature selection problems for common tasks in recommender systems and information retrieval. Additionally, recent studies have explored the use of quantum annealers for instance selection in fine-tuning large models, aiming to optimize training time~\cite{cunha2023effective, pasin2024quantum}. 

\subsubsection{QUBO-based feature selection methods:}

The focus of existing research on quantum annealing based recommender systems is on formalizing problems into QUBO solutions, with the key challenge being how to construct a coefficient matrix $Q$. In the context of the cold-start problem in recommender systems, CQFS~\cite{nembrini2021feature} first trains a recommendation model based solely on user-item interactions, followed by another model trained using features. The coefficient matrix $Q$ is constructed by comparing the consistency of item similarity between the two models. Although Nembrini et al.~\cite{ferrari2022towards} mainly evaluates the applicability of quantum annealers in ranking and classification tasks, they also proposed MIQUBO, CoQUBO and QUBO-boosting. Specifically, MIQUBO fills the coefficient matrix $Q$ with mutual information and conditional mutual information between features and the classification labels, directing the optimization towards using the fewest features to maximize the dependence of the feature set on the classification task. On the other hand, CoQUBO fills the coefficient matrix with the correlation between features and the classification labels, aiming to maximize the relationship between the selected feature set and the classification task. For QUBO-boosting, it constructs the coefficient matrix $Q$ based on the impact of features on the final classification outcome. Similar to feature selection problems, instance selection using quantum annealers~\cite{pasin2024quantum} involves embedding the relevance of training documents into the coefficient matrix, thereby identifying a smaller, more relevant training sample set for the task.

\subsection{Counterfactual Analysis}

Counterfactual analysis is a causal inference tool that introduces perturbations to the internal structure or input of a model, allowing researchers to study its interpretability~\cite{gedikli2014should, tintarev2015explaining} by observing how the model changes before and after the perturbation. This approach helps model designers understand which factors impact the model, enabling them to design more concise and efficient models, improve the model’s final performance, and enhance downstream user experience~\cite{chen2022measuring,ge2022survey,zhang2014explicit}. For instance, the $\text{CF}^{2}$~\cite{tan2022learning} method combines the strengths of both factual reasoning and counterfactual analysis, accurately identifying the nodes and edges in graph structures that are crucial for the current classification task, thereby improving the classification accuracy of graph neural networks. Piccialli et al.~\cite{piccialli2024supervised} applied counterfactual analysis to a pre-trained black-box model to discover more precise discrete boundaries for feature compression, and further inferred the importance of the model by identifying the number of compression boundaries. In the field of recommender systems, ACCENT~\cite{tran2021counterfactual} was the first to propose a framework that applied counterfactual analysis to neural networks. CountER~\cite{tan2021counterfactual} attempted to introduce perturbations to item attribute scores and user preference scores generated by the model, using counterfactual analysis to determine which factors most significantly affect the model’s recommendations, thereby providing interpretability to users. CauseRec~\cite{zhang2021causerec} sampled data from counterfactual data distributions to replace dispensable and indispensable concepts in the original concept sequence, resulting in more accurate and robust user representations. Additionally, some studies have leveraged counterfactual analysis to address fairness and data sparsity issues in recommender systems. For example, PSF-RS~\cite{zhu2023path} splits sensitive features into two latent vectors and distinguishes between them based on the minimal perturbation principle~\cite{blumer1987occam} in counterfactual analysis, thereby extracting useful information from sensitive features for recommendation. CASR~\cite{wang2021counterfactual}, on the other hand, proposed a data augmentation framework using counterfactual analysis. These methods demonstrate the powerful role of counterfactual analysis as a causal inference tool.

\subsection{Positioning within Classical Feature Selection Paradigms}

To situate PDQUBO within the broader feature selection landscape, we summarize classical feature selection paradigms and clarify the relationship between existing QUBO-based approaches and these categories. 

\subsubsection{Classical Feature Selection Taxonomy}

Feature selection methods are generally categorized into three paradigms: filter, wrapper, and embedded methods. Filter methods select features based on statistical criteria independent of a predictive model, such as mutual information or correlation~\cite{peng2005feature,torkkola2003feature,ferreira2012efficient}. Optimization-based approaches such as LPFS~\cite{guo2022lpfs}, which construct objectives from precomputed statistics without repeated retraining, can also be interpreted as filter-like methods. Embedded methods integrate feature selection into model training through sparsity-inducing regularization or learnable gating mechanisms. Examples in recommender systems include Lasso~\cite{tibshirani1996regression}, GBDT feature importance~\cite{friedman2001greedy}, AutoField~\cite{wang2022autofield}, and SHARK~\cite{zhang2023shark}. Wrapper methods treat the predictive model as a black box and evaluate feature subsets according to downstream performance. Typical examples include Sequential Forward Selection (SFS)~\cite{wang2023single} and permutation importance~\cite{fisher2019all}, which assess feature contribution through performance variation.

\subsubsection{QUBO-based Methods and PDQUBO Positioning}

Existing QUBO-based feature selection approaches can be interpreted within this taxonomy. MIQUBO and CoQUBO~\cite{ferrari2022towards} resemble filter methods, as their coefficient matrices are constructed from statistical dependency measures. QUBO-Boosting can be viewed as a model-informed filter, since it leverages model outputs without iterative subset evaluation. CQFS~\cite{nembrini2021feature} is closer to a wrapper-inspired approach, as it evaluates features via model-level similarity consistency. PDQUBO is conceptually aligned with wrapper methods, as it evaluates feature importance through model performance variation under counterfactual perturbations in a post-training setting.

\subsection{Gaps}

Although prior work spans quantum annealing formulations, classical feature selection paradigms, and counterfactual analysis, a fundamental gap remains in directly aligning discrete QUBO optimization with recommendation performance. Existing QUBO-based feature selection methods primarily focus on constructing the coefficient matrix $Q$ using statistical dependency measures or model-derived signals. While these formulations align the optimization objective with generic notions of feature relevance, they do not explicitly optimize downstream recommendation metrics. For instance, MIQUBO and CoQUBO~\cite{ferrari2022towards} rely on feature–label dependency statistics, and QUBO-Boosting incorporates model outputs without modeling performance variation under feature perturbations. CQFS~\cite{nembrini2021feature} introduces collaborative information but evaluates similarity consistency rather than directly optimizing ranking performance. Meanwhile, counterfactual analysis in recommender systems has largely been studied for interpretability, fairness, or robustness, rather than for discrete combinatorial feature selection. As a result, there remains no unified framework that connects counterfactual performance variation with QUBO-based global optimization. To address this gap, we propose PDQUBO, which leverages counterfactual performance perturbations to construct the QUBO coefficient matrix. By explicitly modeling first-order and second-order performance variations, PDQUBO aligns discrete combinatorial optimization with recommendation metrics in a post-training setting.

\section{PRELIMINARIES}
\label{sec:PRL}

\subsection{Notation}
Assume that we have a dataset $\mathcal{D} = (\mathcal{U},\mathcal{V},\mathcal{F})$ used to train a recommender system, where $\mathcal{U} = \{u_1, u_2, \ldots, u_m\}$, $\mathcal{V} = \{v_1, v_2, \ldots, v_n\}$, and $\mathcal{F} = \{f_1, f_2, \ldots, f_{|\mathcal{F}|}\}$ represent the sets of users, items, and item-level features, respectively. Each item $v_i$ is associated with a feature vector $\mathbf{f}_{i} \in \mathbb{R}^{|\mathcal{F}|}$, where each element corresponds to the value of a specific feature in $\mathcal{F}$. We define a binary matrix $B \in\{0,1\}^{m \times n}$ to represent the user-item interaction matrix, where $B_{ij} = 1$ indicates that user $u_i$ has interacted with item $v_j$, and $B_{ij} = 0$ otherwise. The base recommendation model is denoted as $G_{\Theta}$, where $\Theta$ represents the model parameters trained on dataset $\mathcal{D}$. In this work, we use $G(\mathcal{F}|\Theta)_{\text{Mtc}}$ to represent the performance of the trained model $G_{\Theta}$ when evaluated using the full feature set $\mathcal{F}$ under a specific evaluation metric $\text{Mtc}$ (e.g. nDCG~\cite{jarvelin2002cumulated}, or Recall~\cite{herlocker2004evaluating}).

\subsection{QUBO}
Quadratic Unconstrained Binary Optimization (QUBO) is a widely used optimization formulation in quantum computing (e.g., Quantum Annealers)~\cite{glover2018tutorial}, combinatorial optimization (e.g., max-cut, graph coloring), and machine learning~\cite{date2021qubo}. In its standard form, a QUBO problem seeks to minimize a quadratic objective over binary variables:
\begin{equation}
\label{eq:qubo_standard}
\min_{\mathbf{x} \in \{0,1\}^n} Y = \mathbf{x}^T \mathbf{Q} \mathbf{x},
\end{equation}
where $\mathbf{x}$ is an $n$-dimensional binary vector and $\mathbf{Q} \in \mathbb{R}^{n \times n}$ is a symmetric coefficient matrix. The diagonal elements $Q_{ii}$ capture the individual contribution of each variable, while the off-diagonal elements $Q_{ij}$ encode pairwise interactions between variables. 

In the context of feature selection, each binary variable $x_i$ corresponds to a feature $f_i$. Specifically, $x_i = 1$ indicates that feature $f_i$ is selected, and $x_i = 0$ otherwise. The coefficient matrix $\mathbf{Q}$ therefore determines how individual features and their pairwise relationships contribute to the optimization objective, guiding the search toward desirable feature subsets. For practical feature selection tasks, it is often necessary to control the number of selected features. To encourage selecting a target number $k$ of features, we incorporate a soft cardinality constraint through a quadratic penalty term:
\begin{equation}
\label{eq:Equation2}
Y = \mathbf{x}^T \mathbf{Q} \mathbf{x} 
+ \lambda \left( \sum_{i=1}^{|\mathcal{F}|} x_i - k \right)^2,
\end{equation}
where $|\mathcal{F}|$ denotes the total number of candidate features and $k$ is the desired number of selected features. This penalty term is not inherent to the general QUBO formulation, but is introduced here as a modeling choice specific to the feature selection task.

The penalty weight $\lambda$ is set to ensure that the constraint exerts sufficient influence on the optimization process. Following common practice~\cite{ferrari2022towards}, we define:
\begin{equation}
\label{eq:lambda_def}
\lambda = \max_i \sum_j |Q_{ij}|,
\end{equation}
which reflects the maximum possible energy variation contributed by any single variable in the QUBO formulation.

\subsection{Coefficient Matrix $Q$}
\label{sec:QUBO}

The values of the coefficient matrix $Q$ represent the relationship between features and the optimization problem to solve.
For the feature selection problem in recommender systems, let \( \text{Indiv}(f_i) \) represent the contribution of feature \( f_i \) to the performance of the recommendation model, while \( \text{Comb}(f_i, f_j) \) indicate the contribution to recommendation performance when both features \( f_i \) and \( f_j \) are selected together. The optimization goal is to maximize the contribution of the selected feature set to the model's performance. So, $Q$ can be formulated as: 
\begin{equation}
Q_{ij} = \begin{cases}
-\text{Comb}(f_i, f_j) & \text{if } i \neq j \\
-\text{Indiv}(f_i) & \text{if } i = j.
\end{cases}
\label{eq:Equationb}
\end{equation}

\section{Performance Driven QUBO}
\label{sec:METHOD}

\begin{figure}[h]
    \centering
    \includegraphics[width=1.0\linewidth]{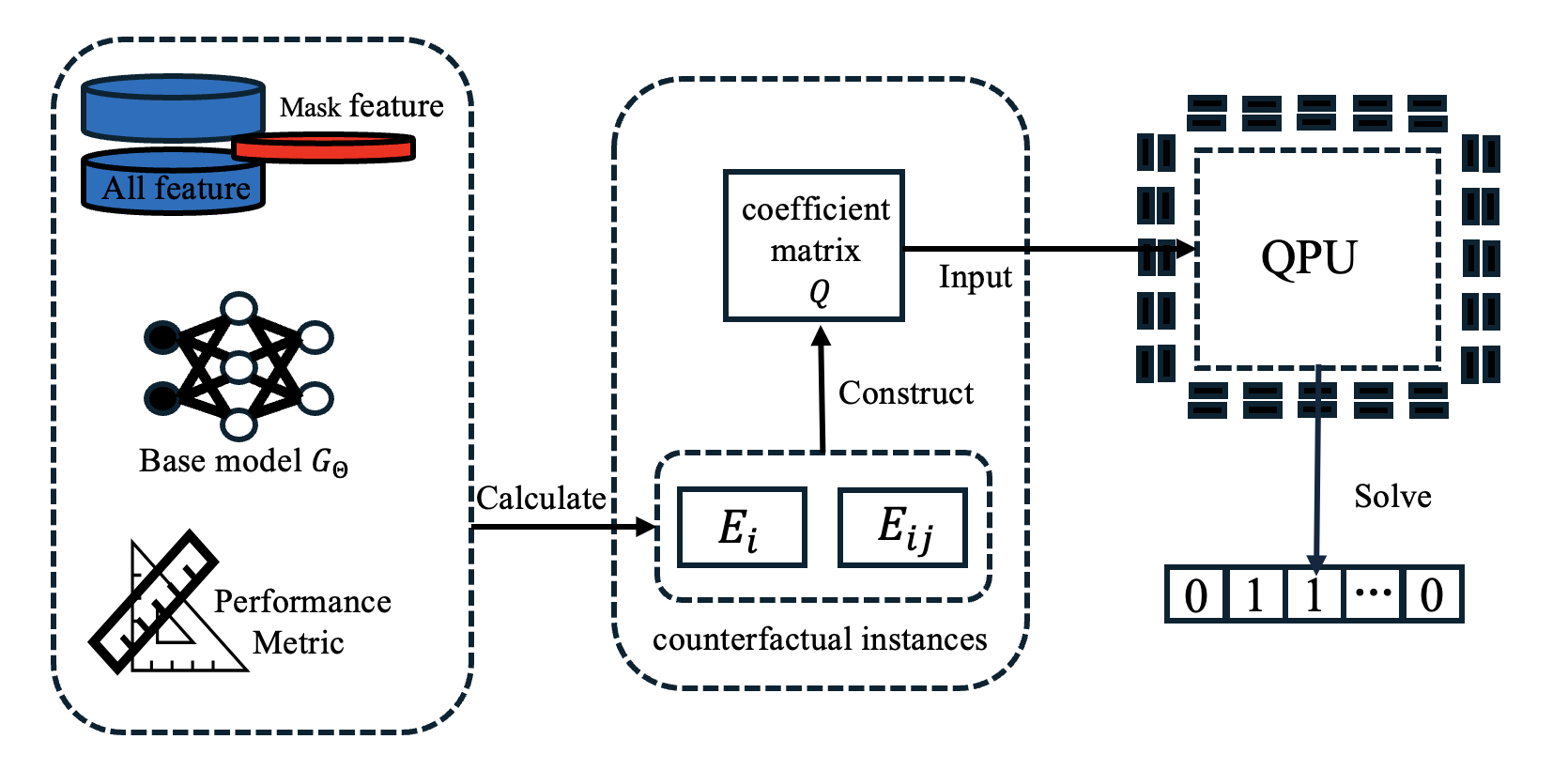}
    \caption{The Overview of PDQUBO}
    \label{fig:PDQUBO}
\end{figure}

We present the \textbf{PDQUBO} model (the overview as shown in Figure~\ref{fig:PDQUBO}) to solve QUBO problems for feature selection in recommender systems.
Then, we detail how \emph{counterfactual instances} are defined, and how they are used to construct the coefficient matrix $Q$.

\subsection{Counterfactual Instances}

Counterfactual Analysis adds perturbations to the base model's input variables and observes the changes before and after the perturbations~\cite{liu2022practical, zhang2021causerec, tan2021counterfactual, wang2024counterfactual}. In this paper, we refer to these changes as \textbf{counterfactual instances}. 
Specifically, similar to~\cite{tan2021counterfactual,zhang2021causerec}, in the context of feature selection for recommender systems, we measure the impact of item features by excluding the corresponding feature and analyzing the difference in recommendation performance between the recommendation lists generated by the base model with and without the corresponding feature.
Following~\cite{chen2024fairgap, tan2022learning, wang2024reinforced}, we define these perturbations in the form of a mask:
\begin{equation}
\mathcal{F_{\text{mask}}} = \mathcal{F} \odot M_{c},
\end{equation}
where $M_{c}$ sets certain features of all items to zero. 
Note, as defined in Equation~\ref{eq:Equation2}, QUBO solves which features to be selected, so $\mathcal{F_{\text{mask}}}$ is defined at feature level accordingly. Namely, these excluded features of all items will be the same.

Next, we employ the recommendation performance metric $Mtc$ (e.g. nDCG~\cite{jarvelin2002cumulated}, or Recall~\cite{herlocker2004evaluating}) for Counterfactual Analysis, which is defined as:
\begin{equation}
\label{eq:Equationd}
\left\{\begin{matrix}
\text{E}_i = G(\mathcal{F}|\Theta)_{\text{Mtc}} - G(\mathcal{F_{\text{mask}}^{\text{i}}}|\Theta)_{\text{Mtc}} \\
\text{E}_{ij} = G(\mathcal{F}|\Theta)_{\text{Mtc}} - G(\mathcal{F_{\text{mask}}^{\text{ij}}}|\Theta)_{\text{Mtc}},
\end{matrix}\right.
\end{equation}
where $E_i$ represents the counterfactual instance of removing $f_i$, and $E_{ij}$ represents the counterfactual instance of removing both $f_i$ and $f_j$. $G(\mathcal{F}|\Theta)_{\text{Mtc}}$ represents the $Mtc$ value obtained by the $G_{\Theta }$ using all item features set $\mathcal{F}$, while $G(\mathcal{F_{\text{mask}}^{\text{i}}}|\Theta)_{\text{Mtc}}$ represents the $Mtc$ value obtained by the $G_{\Theta }$ using features set which is set $\mathcal{F}$ removing $f_i$, The same applies to $G(\mathcal{F_{\text{mask}}^{\text{ij}}}|\Theta)_{\text{Mtc}}$.
It is worth noting that when $E > 0$, it indicates that the performance of the base model decreases after removing the feature, while $E<0$ indicates the performance of the base model improves. While $E = 0$, it means there is no influence on the performance of the base mode when removing corresponding feature(s).

\subsection{Construction of $Q$ for QUBO}
Since the QUBO problem is a minimization problem, we define $Q$ as follows with counterfactual instances $E$:
\begin{equation}
Q_{ij} = 
\begin{cases}
    -E_i,  & \text{if } i = j \\
    -E_{ij}, & \text{if } i \ne j
\end{cases}
\label{eq:cf_qubo}
\end{equation}

Note that $Q_{ii}$ considers only one feature $f_i$, while $Q_{ij}$ considers the pair of features, $f_i$ and $f_j$. This indicates that while the values of $Q$ matter, how many features are considered to obtain those values also matter. Namely, the corresponding counterfactual instances in $Q_{ij}$ when $i \ne j$ would contain more information about features for feature selection purposes. This will be discussed and evaluated in the experiment Sec.~\ref{sec-exp-quadraticterm}.

After constructing $Q$, the feature selection will be solved as a QUBO problem to obtain the final selected features in $\mathbf{x}$, and the corresponding final selected feature set ($\mathcal{F}^*$) will be used for the recommender systems.

\subsection{Quantum Annealing Workflow}

The overall workflow of PDQUBO is summarized as follows. First, a base recommendation model is trained using the complete feature set. After convergence, counterfactual masking is performed to compute the performance variations $E_i$ and $E_{ij}$, which are then used to construct the QUBO coefficient matrix $Q$. Once $Q$ is constructed, no further interaction with the base model is required. The matrix $Q$ is directly submitted to a solver—either classical algorithms (e.g., Simulated Annealing or Tabu Search) or a quantum annealer through the provided API interface. Since QUBO solvers operate solely on the coefficient matrix, the optimization stage is fully decoupled from model training. The solver returns a binary selection vector, which determines the final feature subset used to retrain the base model for evaluation.

In terms of computational cost, the dominant overhead of PDQUBO lies in constructing the matrix $Q$, which requires masked inference using the pre-trained base model. Specifically, computing all counterfactual terms involves $F$ single-feature evaluations and $\frac{F(F-1)}{2}$ pairwise evaluations, where $F$ denotes the number of candidate features. This matrix construction phase is performed offline on classical hardware. Once $Q$ is obtained, the solving stage—whether classical or quantum—only requires the completed coefficient matrix as input. When executed on a QPU, the matrix is submitted directly via the vendor API, which internally manages embedding and annealing procedures. From the user perspective, deploying PDQUBO on a quantum annealer therefore amounts to preparing the QUBO matrix and invoking the solver interface.

\section{EXPERIMENT}{\label{sec:EXP}}
We compare our method with existing QUBO-based feature selection algorithms under various constraint settings, on both classical computers and quantum devices(Sec.~\ref{sec:dataset} and Sec.~\ref{sec-expresults}). Then, we further compare with traditional feature selection methods in the CTR (Click-Through Rate) prediction Task (Sec.~\ref{sec:Ctfsm}).

\subsection{Experimental Design}
\label{sec:dataset}

\subsubsection{Datasets:}

We utilized two datasets, \textbf{150\_ICM } and \textbf{500\_ICM}, provided by CLEF 2024’s QuantumCLEF Lab\footnote{https://qclef.dei.unipd.it/clef2024-lab}, which focuses on benchmarking quantum annealing for information retrieval and recommender systems. The 150\_ICM dataset contains 150 sparse features for each item, while the 500\_ICM dataset contains 500 sparse features for each item. Both datasets are derived from real-world user–item interaction data as described in~\cite{pasin2024overview, niu2024cruise}. In addition to the QuantumCLEF datasets, we further evaluate PDQUBO on two large-scale industrial recommendation datasets (KuaiRec~\cite{gao2022kuairec} and KuaiRand~\cite{gao2022kuairand}) under a fully retrained setting, as detailed in Sec.~\ref{sec:retrain_eval}.

Followed by~\cite{ferrari2022towards}, an 80:20 split is applied to the data to construct the training and test sets. The training set is further divided using another 80:20 split, where 80\% is used to train our PDQUBO model, and the remaining 20\% is reserved for validation purposes. The final results are evaluated on the test set, with all reported metrics representing the average of 5 runs with random initialization.

\subsubsection{Baseline Models:}
To benchmark the performance of our proposed PDQUBO model, we selected a set of well-established baseline models, including CQFS~\cite{nembrini2021feature}, MIQUBO, CoQUBO, and QUBO-Boosting~\cite{ferrari2022towards}:
\begin{itemize}
   \item $~i)$ CQFS: In our implementation, CQFS is constructed using the same base model adopted for PDQUBO (e.g., KNN, MLP-DP, NCF, or MLP-CON). For each dataset, we first train an ID-only version of the base model using only user and item IDs, while removing all item-side content features. The learned latent similarity structure among items is then used as the collaborative signal. A corresponding similarity matrix is computed from the raw feature space, and the alignment between the two similarity structures is incorporated into the QUBO coefficient matrix following the CQFS framework. All models are trained under the same data split and evaluation protocol as PDQUBO.
   \item $~ii)$ MIQUBO and CoQUBO: Following~\cite{ferrari2022towards},MIQUBO fills the coefficient matrix with mutual information and conditional mutual information between features and labels, while CoQUBO fills the matrix with the correlation between features and labels. The classification labels are replaced with interactions in the recommender system, and interaction data undergoes 1:1 negative sampling.
   \item $~iii)$ QUBO-Boosting: Following \cite{ferrari2022towards}, QUBO-Boosting fills the coefficient matrix with the predicted values from the Support Vector Classifier and the true labels. The SVC is replaced with the base models.
\end{itemize}

\subsubsection{Metrics:}
We evaluated model performance using the Normalized Discounted Cumulative Gain (nDCG), a metric that measures ranking quality by considering both the relevance of items and their positions within the top-$N$ recommendations. This is particularly effective for recommender systems, as it takes into account not only the presence of relevant items but also their order in the ranked list.
We set $N$ to 10 in this study.

\subsubsection{Base Models:}
To thoroughly evaluate the performance of the proposed PDQUBO model across various base recommendation models, we tested approaches ranging from classical methods to neural network-based models: 
\begin{itemize}
    \item $~i)$ \emph{Item-KNN~\cite{sarwar2001item}}, a classical model that uses the user-item interaction matrix to predict potential interactions based on similarities calculated using item features; 
    \item $~ii)$ \emph{MLP-DP/MLP-CON}, we developed two neural network-based models: MLP-DP and MLP-CON, both utilizing a Multi-Layer Perceptron (MLP~\cite{da2020recommendation}), to evaluate PDQUBO's performance with different item feature processing strategies. These models allow us to investigate various methods for integrating item features. MLP-DP fuses the item’s latent embedding with its features, passing this combined data through an MLP. The result is then used in a dot product with the user’s embedding to predict interactions. MLP-CON, on the other hand, concatenates the item and user embeddings, along with the item's features, into a single vector, which is passed through an MLP to generate predictions. Both models employ Bayesian Personalized Ranking (BPR~\cite{rendle2012bpr}) loss with a 1:1 negative-to-positive sample ratio for optimization. These experiments provide valuable insights into how different feature processing techniques impact model prediction performance; 
    \item $~iii)$ \emph{NCF~\cite{he2017neural}}, Neural Collaborative Filtering (NCF) leverages both an MLP and Generalized Matrix Factorization (GMF) to model the non-linear and linear interactions between users and items. In this paper, we feed item features into an MLP to produce outputs matching the item embeddings' dimensionality, which are then summed with the embeddings to integrate the features. This directly incorporates item features into the NCF architecture. Like MLP-DP and MLP-CON, we optimize NCF with BPR loss, maintaining a 1:1 ratio of negative to positive samples during training.
\end{itemize}

\subsubsection{Hyperparameter Tuning:}
For Item-KNN, we utilized a fixed number of neighbors (we set it as 100 by following CLEF 2024's QuantumCLEF Lab) to calculate predictions. For the neural network-based base models, hyperparameters were first tuned via grid search~\cite{mantovani2015effectiveness} on the validation set prior to feature selection. Since PDQUBO relies on counterfactual analysis derived from a trained base model, it is necessary to establish a stable and reasonably optimized decision function before constructing the QUBO matrix. Grid search provides a deterministic and reproducible hyperparameter selection protocol, ensuring that all feature subsets are evaluated under the same trained model configuration in the main experiments. In the additional fully retrained experiments (Sec.~\ref{sec:retrain_eval}), we adopt Bayesian optimization to re-tune hyperparameters after feature selection. This adaptive setting reflects practical deployment scenarios and verifies that the observed performance gains are not artifacts of fixed-parameter evaluation. The detailed search configurations for both grid search and Bayesian optimization are summarized in Table~\ref{tab:hyper_search}.

\begin{table}[t]
\centering
\caption{Hyperparameter search configurations.}
\label{tab:hyper_search}
\footnotesize
\resizebox{0.85\linewidth}{!}{
\begin{tblr}{
  width = 0.9\linewidth,
  colspec = {Q[100]Q[150]Q[250]},
  cells = {c},
  hline{1,6} = {-}{0.15em},
  hline{2} = {1-3}{0.15em},
  vline{1,4} = {-}{0.15em},
}
\textbf{Hyperparameter} 
& \textbf{Grid Search} 
& \textbf{Bayesian Optimization} \\

Embedding dimension 
& \{64, 128\} 
& Integer: $[64, 256]$ \\

MLP hidden layers 
& \{\texttt{None}, \texttt{128\_64}, \texttt{256\_128}\} 
& Categorical: \{\texttt{None}, \texttt{64\_32}, \texttt{128\_64}, \texttt{128\_32}, 
\texttt{256\_64}, \texttt{256\_128}, \texttt{256\_128\_32}, \texttt{512\_256\_64}\} \\

Learning rate 
& \{1e{-4}, 3e{-4}, 1e{-3}, 3e{-3}, 1e{-2}, 3e{-2}\} 
& Real: $[10^{-4}, 10^{-2}]$ (log-uniform) \\

Weight decay 
& \{0, 1e{-6}, 1e{-4}\} 
& Real: $[10^{-6}, 10^{-3}]$ (log-uniform) \\
\end{tblr}
}
\end{table}

\subsubsection{The Configuration of $k$:}
The values of the number of features selected, $k$ (in Equation~\ref{eq:Equation2}) are set to [130, 135, 140, 145] for 150\_ICM, and [350, 400, 450, 470] for 500\_ICM. Additionally, we allow both PDQUBO and all baseline models to automatically determine the optimal number of features to maximize recommendation performance, indicated by $*$.

\subsubsection{QUBO Optimization Methods:}
To assess the performance of traditional versus quantum optimization techniques, we selected traditional methods that can be used to optimize QUBO problems, including Simulated Annealing (SA)\cite{bertsimas1993simulated}, Stochastic Gradient Descent (SGD)\cite{zinkevich2010parallelized}, and Tabu Search (TS)\cite{glover1990tabu} as comparison methods. These are compared against Quantum Annealing (QA)\cite{rajak2023quantum} and Hybrid~\cite{McGeoch2020TheDA} approaches implemented on D-Wave\footnote{Access to the D-Wave quantum annealer can be obtained through the Leap cloud platform: \url{https://cloud.dwavesys.com/leap/login?next=/leap}}~\cite{McGeoch2020TheDA}, both of which are Quantum Processing Unit (QPU)-based. Although the D-Wave Advantage 6.1 system provides over 5,000 physical qubits, the sparse connectivity of its Pegasus topology and the overhead of minor-embedding significantly reduce the number of logical variables that can be represented, often to below a few hundred for dense QUBO problems. As a result, QA alone is not suitable for directly solving large-scale problems with hundreds or more variables. The Hybrid method, which integrates classical and quantum optimization, can overcome this limitation and is particularly effective for handling data sizes that exceed the practical qubit capacity~\cite{devoret2004superconducting}. The implementation of our method is available at https://github.com/jiayangniu/PDQUBO.git

\subsection{Experimental Results}
\label{sec-expresults}

\begin{table}[!t]
\centering
\caption{The performance of nDCG@10 using different optimization methods on various datasets under different base models. The bold text represents the best result of the five feature selection methods under the same conditions. The rows labeled \texttt{150\_ICM} and \texttt{500\_ICM} correspond to the performance of the base models without any feature selection. The upper part of the table reports the results on the \texttt{150\_ICM} dataset, while the lower part reports the results on the \texttt{500\_ICM} dataset. Note that for \texttt{500\_ICM}, the QA method on QPU could not be executed due to hardware limitations, and thus is indicated with ``-''.}
\label{tab:table1}
\resizebox{1.0\linewidth}{!}{
\tiny
\begin{tblr}{
  width = \linewidth,
  colspec = {Q[20]Q[295]Q[35]Q[35]Q[35]Q[35]Q[35]Q[35]Q[35]Q[35]Q[35]Q[35]Q[35]Q[35]Q[35]Q[35]Q[35]Q[35]Q[35]Q[35]Q[35]Q[35]},
  cells = {c},
  cell{1}{1} = {c=2}{0.093\linewidth},
  cell{1}{3} = {c=5}{0.202\linewidth},
  cell{1}{8} = {c=5}{0.202\linewidth},
  cell{1}{13} = {c=5}{0.202\linewidth},
  cell{1}{18} = {c=5}{0.202\linewidth},
  cell{2}{1} = {r=3}{},
  cell{2}{2} = {r=3}{},
  cell{2}{3} = {c=3}{0.12\linewidth},
  cell{2}{6} = {c=2}{0.082\linewidth},
  cell{2}{8} = {c=3}{0.12\linewidth},
  cell{2}{11} = {c=2}{0.082\linewidth},
  cell{2}{13} = {c=3}{0.12\linewidth},
  cell{2}{16} = {c=2}{0.082\linewidth},
  cell{2}{18} = {c=3}{0.12\linewidth},
  cell{2}{21} = {c=2}{0.082\linewidth},
  cell{5}{1} = {c=2}{0.093\linewidth},
  cell{5}{3} = {c=5}{0.202\linewidth},
  cell{5}{8} = {c=5}{0.202\linewidth},
  cell{5}{13} = {c=5}{0.202\linewidth},
  cell{5}{18} = {c=5}{0.202\linewidth},
  cell{6}{1} = {r=5}{},
  cell{11}{1} = {r=5}{},
  cell{16}{1} = {r=5}{},
  cell{21}{1} = {r=5}{},
  cell{26}{1} = {r=5}{},
  cell{31}{1} = {c=2}{0.093\linewidth},
  cell{31}{3} = {c=5}{0.202\linewidth},
  cell{31}{8} = {c=5}{0.202\linewidth},
  cell{31}{13} = {c=5}{0.202\linewidth},
  cell{31}{18} = {c=5}{0.202\linewidth},
  cell{32}{1} = {r=5}{},
  cell{37}{1} = {r=5}{},
  cell{42}{1} = {r=5}{},
  cell{47}{1} = {r=5}{},
  cell{52}{1} = {r=5}{},
  vline{1, 3, 8, 13, 18, 23} = {-}{0.2em},
  vline{2} = {2-57}{0.12em},
  vline{6, 11, 16, 21} = {2-4}{0.12em},
  hline{1, 2, 5, 6, 31, 32, 57} = {-}{0.2em},
  hline{11, 16, 21, 26, 37, 42, 47, 52} = {-}{0.12em},
  hline{4} = {3-23}{0.12em},
}
Model                &               & KNN       &      &      &      &        & MLP-DP    &      &      &      &        & NCF       &      &      &      &        & MLP-CON       &      &      &      &        \\
F                    & Mtd.          & Tradition &      &      & QPU  &        & Tradition &      &      & QPU  &        & Tradition &      &      & QPU  &        & Tradition &      &      & QPU  &        \\
                     &               & SA        & SGD  & TS   & QA   & Hybrid & SA        & SGD  & TS   & QA   & Hybrid & SA        & SGD  & TS   & QA   & Hybrid & SA        & SGD  & TS   & QA   & Hybrid \\
                     &               & N@10      & N@10 & N@10 & N@10 & N@10   & N@10      & N@10 & N@10 & N@10 & N@10   & N@10      & N@10 & N@10 & N@10 & N@10   & N@10      & N@10 & N@10 & N@10 & N@10   \\
150\_ICM             &               & 0.1028    &      &      &      &        & 0.1308    &      &      &      &        & 0.1336    &      &      &      &        & 0.1180    &      &      &      &        \\
$*$                  & CQFS          & 0.0112          & 0.0095          & 0.0134          & 0.0495          & 0.0112          & 0.0108          & 0.0078          & 0.0176          & 0.0639          & 0.0108          & 0.0874          & 0.0868          & 0.0874          & 0.0868          & 0.0874          & 0.0042          & 0.0020          & 0.0047          & 0.0710          & 0.1062          \\
                     & QUBO-BOOSTING & 0.1028          & 0.1028          & 0.1028          & 0.1028          & 0.1028          & 0.1308          & 0.1308          & 0.1308          & 0.1308          & 0.1308          & 0.1336          & 0.1336          & 0.1336          & 0.1336          & 0.1336          & 0.1180          & 0.1180          & 0.1180          & 0.1180          & 0.1180          \\
                     & CoQUBO        & 0.0319          & 0.0309          & 0.0323          & 0.0527          & 0.0319          & 0.0356          & 0.0253          & 0.0363          & 0.0486          & 0.0356          & 0.0293          & 0.0204          & 0.0347          & 0.0212          & 0.0293          & 0.0074          & 0.0057          & 0.0074          & 0.0094          & 0.0074          \\
                     & MIQUBO        & 0.1028          & 0.1028          & 0.1028          & 0.1028          & 0.1028          & 0.1308          & 0.1308          & 0.1308          & 0.1308          & \textbf{0.1308} & 0.1336          & 0.1336          & 0.1336          & 0.1336          & 0.1336          & 0.1180          & 0.1180          & 0.1180          & 0.1180          & 0.1180          \\
                     & PDQUBO        & \textbf{0.1168} & \textbf{0.1168} & \textbf{0.1168} & \textbf{0.1168} & \textbf{0.1168} & 0.1299          & 0.1299          & 0.1299          & 0.1299          & 0.1299          & \textbf{0.1366} & \textbf{0.1366} & \textbf{0.1366} & \textbf{0.1366} & \textbf{0.1366} & 0.1116          & 0.1116          & 0.1116          & 0.1116          & 0.1116          \\
130                  & CQFS          & 0.0983          & 0.1029          & 0.0962          & 0.1009          & 0.0955          & 0.1328          & 0.1312          & \textbf{0.1350} & 0.1295          & 0.1343          & 0.1310          & 0.1308          & 0.1336          & 0.1304          & 0.1322          & 0.1102          & 0.1115          & 0.1144          & 0.1067          & 0.1163          \\
                     & QUBO-BOOSTING & 0.1021          & 0.1039          & 0.0954          & 0.0905          & 0.0952          & 0.1318          & 0.1314          & 0.1350          & 0.1288          & 0.1334          & 0.1301          & 0.1308          & 0.1305          & 0.1272          & 0.1305          & 0.1167          & 0.1089          & 0.1130          & 0.1142          & 0.1147          \\
                     & CoQUBO        & 0.0948          & 0.1033          & 0.0941          & 0.0898          & 0.0946          & 0.1332          & 0.1320          & 0.1295          & 0.1332          & 0.1305          & 0.1263          & 0.1309          & 0.1259          & 0.1263          & 0.1244          & 0.1078          & 0.1105          & 0.0980          & 0.1102          & 0.1035          \\
                     & MIQUBO        & 0.1018          & \textbf{0.1040} & 0.1022          & \textbf{0.1072} & 0.1033          & 0.1276          & 0.1307          & 0.1305          & 0.1335          & 0.1273          & 0.1314          & 0.1318          & 0.1315          & 0.1256          & 0.1300          & 0.1106          & 0.1138          & 0.1160          & 0.0938          & 0.1103          \\
                     & PDQUBO        & \textbf{0.1140} & 0.1025          & \textbf{0.1168} & 0.0877          & \textbf{0.1151} & \textbf{0.1366} & \textbf{0.1331} & 0.1308          & \textbf{0.1354} & \textbf{0.1346} & \textbf{0.1343} & 0.1307          & \textbf{0.1363} & 0.1307          & \textbf{0.1347} & 0.1156          & 0.1106          & 0.1153          & 0.1117          & 0.1132          \\
135                  & CQFS          & 0.0996          & 0.1035          & 0.0981          & 0.1008          & 0.1095          & 0.1320          & 0.1310          & 0.1342          & 0.1280          & 0.1321          & 0.1318          & 0.1306          & 0.1329          & 0.1251          & 0.1333          & 0.1133          & 0.1107          & 0.1123          & 0.0954          & 0.1149          \\
                     & QUBO-BOOSTING & 0.0968          & 0.1058          & 0.0969          & 0.0979          & 0.1012          & 0.1330          & 0.1306          & \textbf{0.1360} & 0.1311          & 0.1341          & 0.1310          & 0.1308          & 0.1296          & 0.1308          & 0.1324          & 0.1160          & 0.1118          & 0.1147          & 0.1129          & 0.1150          \\
                     & CoQUBO        & 0.0973          & 0.1044          & 0.1009          & 0.0919          & 0.0983          & 0.1323          & 0.1308          & 0.1332          & \textbf{0.1335} & 0.1320          & 0.1280          & 0.1305          & 0.1284          & 0.1285          & 0.1280          & 0.1088          & 0.1079          & 0.1081          & 0.1100          & 0.1090          \\
                     & MIQUBO        & 0.1021          & 0.1039          & 0.1018          & 0.0912          & 0.1020          & 0.1282          & 0.1300          & 0.1308          & 0.1255          & 0.1294          & 0.1315          & 0.1303          & 0.1320          & 0.1279          & 0.1323          & 0.1136          & 0.1069          & \textbf{0.1181} & 0.1099          & 0.1144          \\
                     & PDQUBO        & \textbf{0.1126} & 0.1044          & \textbf{0.1173} & 0.0947          & \textbf{0.1137} & \textbf{0.1350} & \textbf{0.1319} & 0.1335          & 0.1299          & \textbf{0.1350} & \textbf{0.1345} & 0.1307          & \textbf{0.1365} & 0.1226          & \textbf{0.1345} & 0.1147          & 0.1089          & 0.1131          & \textbf{0.1188} & 0.1158          \\
140                  & CQFS          & 0.1015          & \textbf{0.1053} & 0.1012          & 0.1051          & 0.1012          & 0.1330          & 0.1315          & 0.1331          & 0.1303          & 0.1331          & 0.1326          & 0.1318          & 0.1337          & 0.1315          & 0.1340          & 0.1167          & 0.1143          & 0.1154          & 0.1143          & 0.1167          \\
                     & QUBO-BOOSTING & 0.0986          & 0.1038          & 0.0987          & 0.0879          & 0.0967          & 0.1325          & 0.1304          & \textbf{0.1350} & \textbf{0.1335} & 0.1350          & 0.1316          & 0.1314          & 0.1324          & 0.1335          & 0.1324          & 0.1179          & 0.1131          & 0.1138          & 0.1168          & 0.1179          \\
                     & CoQUBO        & 0.1007          & 0.1021          & 0.1028          & 0.1060          & 0.1028          & 0.1335          & 0.1309          & 0.1316          & 0.1333          & 0.1316          & 0.1307          & 0.1314          & 0.1322          & 0.1328          & 0.1322          & 0.1112          & 0.1128          & 0.1176          & 0.1137          & 0.1176          \\
                     & MIQUBO        & 0.1028          & 0.1038          & 0.1032          & 0.1006          & 0.1027          & 0.1295          & 0.1305          & 0.1306          & 0.1307          & 0.1297          & 0.1320          & 0.1311          & 0.1333          & 0.1322          & 0.1324          & 0.1137          & 0.1119          & 0.1183          & 0.1137          & 0.1166          \\
                     & PDQUBO        & \textbf{0.1121} & 0.1028          & \textbf{0.1191} & 0.1054          & \textbf{0.1135} & \textbf{0.1356} & \textbf{0.1318} & 0.1350          & 0.1293          & \textbf{0.1355} & \textbf{0.1339} & 0.1318          & \textbf{0.1363} & 0.1319          & \textbf{0.1350} & 0.1178          & 0.1156          & \textbf{0.1193} & 0.1140          & \textbf{0.1193} \\
145                  & CQFS          & 0.1028          & \textbf{0.1036} & 0.1028          & 0.1031          & 0.1028          & 0.1320          & 0.1314          & 0.1320          & 0.1276          & 0.1320          & 0.1338          & 0.1326          & \textbf{0.1345} & 0.1236          & 0.1331          & 0.1180          & 0.1168          & 0.1179          & 0.1164          & 0.1185          \\
                     & QUBO-BOOSTING & 0.1019          & 0.1027          & 0.1019          & 0.1056          & 0.1003          & 0.1323          & 0.1307          & \textbf{0.1330} & 0.1268          & \textbf{0.1323} & 0.1339          & 0.1320          & 0.1336          & 0.1264          & 0.1330          & 0.1180          & 0.1146          & 0.1166          & 0.1129          & 0.1178          \\
                     & CoQUBO        & 0.1028          & 0.1027          & 0.1030          & 0.0949          & 0.1030          & 0.1314          & 0.1312          & 0.1317          & \textbf{0.1334} & 0.1317          & 0.1325          & 0.1326          & 0.1333          & 0.1310          & 0.1333          & 0.1159          & 0.1155          & 0.1165          & 0.1135          & 0.1165          \\
                     & MIQUBO        & 0.1031          & 0.1035          & 0.1030          & 0.1011          & 0.1028          & 0.1304          & 0.1313          & 0.1306          & 0.1307          & 0.1309          & 0.1331          & 0.1328          & 0.1334          & 0.1316          & 0.1332          & 0.1179          & 0.1158          & 0.1184          & 0.1166          & 0.1178          \\
                     & PDQUBO        & \textbf{0.1125} & 0.1029          & \textbf{0.1116} & \textbf{0.1061} & \textbf{0.1116} & \textbf{0.1334} & \textbf{0.1316} & 0.1322          & 0.1284          & 0.1322          & \textbf{0.1345} & 0.1326          & 0.1338          & \textbf{0.1336} & \textbf{0.1338} & \textbf{0.1205} & 0.1155          & \textbf{0.1203} & 0.1050          & \textbf{0.1203} \\
500\_ICM             &               & 0.1092          &                 &                 &                 &                 & 0.1401          &                 &                 &                 &                 & 0.1299          &                 &                 &                 &                 & 0.1087          &                 &                 &                 &                 \\
$*$                  & CQFS          & 0.0034          & 0.0308          & 0.0275          & -               & 0.0308          & 0.0001          & 0.0196          & 0.0209          & -               & 0.0081          & 0.0706          & 0.1157          & 0.1157          & -               & 0.1157          & 0.0033          & 0.0081          & 0.0088          & -               & 0.0069          \\
                     & QUBO-BOOSTING & 0.1092          & 0.1092          & 0.1092          & -               & 0.1092          & 0.1401          & 0.1401          & 0.1401          & -               & 0.1401          & 0.1299          & 0.1299          & 0.1299          & -               & 0.1299          & 0.1087          & 0.1087          & 0.1087          & -               & 0.1087          \\
                     & CoQUBO        & 0.0308          & 0.0436          & 0.0352          & -               & 0.0309          & 0.0185          & 0.0278          & 0.0194          & -               & 0.0180          & 0.0101          & 0.0113          & 0.0112          & -               & 0.0092          & 0.0973          & 0.0151          & 0.0094          & -               & 0.0078          \\
                     & MIQUBO        & 0.1092          & 0.1092          & 0.1092          & -               & 0.1092          & 0.1401          & 0.1401          & 0.1401          & -               & 0.1401          & 0.1299          & 0.1299          & 0.1299          & -               & 0.1299          & 0.1087          & 0.1087          & 0.1087          & -               & 0.1087          \\
                     & PDQUBO        & \textbf{0.1359} & \textbf{0.1359} & \textbf{0.1359} & -               & \textbf{0.1359} & 0.1365          & 0.1365          & 0.1365          & -               & 0.1365          & \textbf{0.1306} & \textbf{0.1306} & \textbf{0.1306} & -               & \textbf{0.1306} & 0.1036          & 0.1036          & 0.1036          & -               & 0.1036          \\
350                  & CQFS          & 0.1033          & 0.1093          & 0.1067          & -               & 0.1090          & 0.1397          & 0.1382          & 0.1220          & -               & 0.1390          & 0.1219          & 0.1299          & 0.1183          & -               & 0.1200          & 0.0942          & 0.1019          & 0.0842          & -               & 0.0998          \\
                     & QUBO-BOOSTING & 0.0955          & 0.1144          & 0.0832          & -               & 0.0897          & 0.1386          & 0.1375          & 0.1398          & -               & 0.1371          & 0.1214          & 0.1291          & 0.1207          & -               & 0.1174          & 0.0969          & 0.1052          & 0.0942          & -               & 0.0938          \\
                     & CoQUBO        & 0.0866          & 0.1153          & 0.0541          & -               & 0.0693          & 0.1358          & 0.1387          & 0.1222          & -               & 0.1157          & 0.1138          & 0.1295          & 0.0961          & -               & 0.1030          & 0.0993          & 0.1115          & 0.0620          & -               & 0.0736          \\
                     & MIQUBO        & 0.1036          & 0.1101          & 0.1033          & -               & 0.1069          & 0.1380          & 0.1374          & 0.1373          & -               & 0.1355          & 0.1255          & 0.1316          & 0.1287          & -               & 0.1251          & 0.0962          & 0.1036          & \textbf{0.1098} & -               & 0.1022          \\
                     & PDQUBO        & \textbf{0.1279} & 0.1093          & \textbf{0.1356} & -               & \textbf{0.1289} & 0.1393          & 0.1387          & 0.1366          & -               & 0.1391          & 0.1288          & 0.1292          & 0.1280          & -               & \textbf{0.1335} & 0.1070          & 0.1098          & 0.1085          & -               & \textbf{0.1105} \\
400                  & CQFS          & 0.1074          & \textbf{0.1096} & 0.1056          & -               & 0.1059          & 0.1408          & 0.1371          & 0.1331          & -               & 0.1402          & 0.1274          & 0.1307          & 0.1201          & -               & 0.1245          & 0.1037          & 0.1059          & 0.0932          & -               & 0.0975          \\
                     & QUBO-BOOSTING & 0.0926          & 0.1093          & 0.0861          & -               & 0.0937          & 0.1420          & 0.1381          & \textbf{0.1413} & -               & 0.1403          & 0.1240          & 0.1298          & 0.1233          & -               & 0.1240          & 0.0990          & 0.1078          & 0.0951          & -               & 0.0996          \\
                     & CoQUBO        & 0.0922          & 0.1079          & 0.0630          & -               & 0.0749          & 0.1405          & 0.1397          & 0.1321          & -               & 0.1371          & 0.1268          & 0.1314          & 0.1140          & -               & 0.1139          & 0.0999          & 0.1084          & 0.0714          & -               & 0.0846          \\
                     & MIQUBO        & 0.1077          & 0.1093          & 0.1052          & -               & 0.1150          & 0.1384          & 0.1394          & 0.1373          & -               & 0.1368          & 0.1286          & 0.1315          & \textbf{0.1306} & -               & 0.1259          & 0.1067          & 0.1098          & \textbf{0.1106} & -               & 0.1074          \\
                     & PDQUBO        & \textbf{0.1320} & 0.1091          & \textbf{0.1354} & -               & \textbf{0.1370} & \textbf{0.1430} & 0.1391          & 0.1383          & -               & \textbf{0.1441} & 0.1299          & 0.1294          & 0.1291          & -               & \textbf{0.1331} & \textbf{0.1093} & \textbf{0.1127} & 0.1095          & -               & \textbf{0.1127} \\
450                  & CQFS          & 0.1120          & \textbf{0.1102} & 0.1163          & -               & 0.1102          & 0.1406          & 0.1401          & 0.1429          & -               & 0.1415          & 0.1295          & 0.1305          & 0.1269          & -               & 0.1268          & 0.1085          & 0.1069          & 0.1065          & -               & 0.1073          \\
                     & QUBO-BOOSTING & 0.0987          & 0.1076          & 0.0938          & -               & 0.0934          & 0.1413          & \textbf{0.1407} & 0.1423          & -               & \textbf{0.1431} & 0.1255          & 0.1299          & 0.1269          & -               & 0.1268          & 0.1023          & 0.1078          & 0.1006          & -               & 0.1006          \\
                     & CoQUBO        & 0.1026          & 0.1079          & 0.0967          & -               & 0.0923          & 0.1413          & 0.1407          & 0.1413          & -               & 0.1412          & 0.1283          & 0.1320          & 0.1259          & -               & 0.1264          & 0.1010          & 0.1090          & 0.0960          & -               & 0.0948          \\
                     & MIQUBO        & 0.1097          & 0.1089          & 0.1087          & -               & 0.1086          & 0.1395          & 0.1399          & 0.1392          & -               & 0.1377          & \textbf{0.1324} & 0.1301          & 0.1302          & -               & 0.1279          & \textbf{0.1118} & 0.1088          & 0.1091          & -               & 0.1044          \\
                     & PDQUBO        & \textbf{0.1339} & 0.1083          & \textbf{0.1426} & -               & \textbf{0.1370} & \textbf{0.1424} & 0.1404          & \textbf{0.1432} & -               & 0.1424          & \textbf{0.1324} & \textbf{0.1340} & \textbf{0.1314} & -               & \textbf{0.1329} & 0.1107          & \textbf{0.1115} & \textbf{0.1103} & -               & \textbf{0.1128} \\
470                  & CQFS          & 0.1128          & \textbf{0.1143} & 0.1155          & -               & 0.1143          & 0.1401          & 0.1401          & 0.1411          & -               & 0.1403          & 0.1303          & 0.1304          & 0.1282          & -               & 0.1267          & 0.1101          & 0.1075          & \textbf{0.1104} & -               & 0.1091          \\
                     & QUBO-BOOSTING & 0.1042          & 0.1076          & 0.0984          & -               & 0.0980          & 0.1411          & 0.1402          & \textbf{0.1450} & -               & \textbf{0.1449} & 0.1290          & 0.1308          & 0.1277          & -               & 0.1278          & 0.1069          & 0.1080          & 0.1022          & -               & 0.1018          \\
                     & CoQUBO        & 0.1017          & 0.1086          & 0.1002          & -               & 0.1017          & 0.1414          & \textbf{0.1408} & 0.1409          & -               & 0.1417          & 0.1296          & 0.1309          & 0.1286          & -               & 0.1271          & 0.1025          & 0.1083          & 0.1034          & -               & 0.0971          \\
                     & MIQUBO        & 0.1092          & 0.1081          & 0.1093          & -               & 0.1085          & 0.1398          & 0.1406          & 0.1396          & -               & 0.1402          & 0.1311          & 0.1313          & 0.1299          & -               & 0.1303          & 0.1093          & 0.1092          & 0.1090          & -               & 0.1078          \\
                     & PDQUBO        & \textbf{0.1373} & 0.1083          & \textbf{0.1379} & -               & \textbf{0.1402} & \textbf{0.1449} & 0.1406          & 0.1416          & -               & 0.1428          & \textbf{0.1326} & \textbf{0.1342} & \textbf{0.1322} & -               & \textbf{0.1329} & \textbf{0.1135} & \textbf{0.1127} & 0.1102          & -               & \textbf{0.1102}       
\end{tblr}
}
\end{table}

\subsubsection{Baselines comparison:}

This section provides a comprehensive analysis of PDQUBO’s performance compared to baseline models (CQFS, QUBO-Boosting, CoQUBO, and MIQUBO) across two datasets (150ICM and 500ICM), different optimization methods (SA, SGD, TS, QA, Hybrid), various number of features ($k$), and base models (Item-KNN, MLP-DP, NCF, MLP-CON). 
In Table~\ref{tab:table1}, the performance after feature selection is obtained by directly evaluating the base models using the parameters from the counterfactual instance generation stage, without retraining or hyperparameter re-optimization. This design is intentional: since the QUBO matrix is constructed from counterfactual analysis of a trained base model, retraining the model after feature selection would entangle the intrinsic effect of feature removal with subsequent parameter adaptation. Therefore, the fixed-parameter evaluation isolates the pure contribution of feature selection methods under a shared model configuration.
We acknowledge that this controlled setting may introduce fluctuations in absolute performance. To address this limitation, Sec.~\ref{sec:retrain_eval} presents complementary experiments in which the base models are retrained and hyperparameters are re-optimized after feature selection, resulting in more stable and deployment-oriented performance patterns. Special attention is given to the case when $k=*$, where the model is allowed to automatically select the optimal number of features. We did not optimize models using the QA optimization method in the 500ICM dataset due to quantum computational limitations, with `-' representing the cells where results are not available. All results of nDCG@10 are shown in Table~\ref{tab:table1}.

In both 150ICM and 500ICM datasets, PDQUBO generally demonstrates strong performance across different optimization methods. 
However, there are scenarios where the baselines perform comparably or even better, depending on the optimization method, base models, and feature selection value. For the SA method, PDQUBO shows notable improvements in many cases. For example, in the Item-KNN model, when $k=130$, the best-performing baseline is QUBO-Boosting with an nDCG@10 of 0.1021. PDQUBO achieves a higher score of 0.1140, reflecting an 11.7\% improvement. Similarly, at $k=140$, the best baseline is CQFS with 0.1015, while PDQUBO delivers 0.1121, marking a 10.4\% gain. However, at $k=145$, the gap narrows as PDQUBO achieves 0.1125, compared to QUBO-Boosting's and CoQUBO's 0.1028. While PDQUBO outperforms the baseline, the improvement is less pronounced at higher $k$ values. In the case of SGD, PDQUBO shows mixed results. For example, in the MLP-DP, at $k=130$, PDQUBO delivers 0.1331, compared to the best baseline CoQUBO with 0.1320, showing a modest 0.83\%  improvement. However, in the Item-KNN, NCF, and MLP-CON, the best-performing baselines nearly match PDQUBO, even better than PDQUBO at different values of $k$. For TS, PDQUBO also demonstrates competitive performance, particularly in the Item-KNN, and NCF at various $k$. For example, in the NCF, at $k=130$, the best baseline is CQFS with an nDCG@10 of 0.1336, while PDQUBO achieves 0.1363, representing a 2.0\% improvement. For QA, PDQUBO's performance is also competitive in some cases. For instance, in the NCF, at $k=*$, PDQUBO achieves 0.1366, a slight improvement over the best baseline MIQUBO, which scores 0.1336. However, when $k$ is set from 130 to 145, PDQUBO struggles to maintain competitive performance. For Hyrid, PDQUBO demonstrates particularly strong performance in Item-KNN, MLP-DP, and NCF. For example, in the NCF, at $k=145$, PDQUBO achieves 0.1338, compared to CQFS's 0.1331, representing a 0.53\% improvement. Similar trends are observed in the 500ICM dataset, with the exception of the QA optimization method, which was not evaluated. Across both datasets 150ICM and 500ICM, PDQUBO demonstrates superior performance over the baseline models in most of the cases. However, in some cases, the baselines perform comparably or even better, especially in SGD and QA optimization methods. However, we note that the fixed-parameter evaluation in Table~\ref{tab:table1} can lead to fluctuations and inconsistencies in performance. Recognizing this limitation, Sec.~\ref{sec:retrain_eval} presents complementary experiments in which the base models are retrained and tuned after feature selection, resulting in more stable performance patterns.

\begin{table}[t]
\centering
\caption{Comparison of energy $Y$ with vs without constrains. ``without constrains" means optimizatin methods freely decide how many featrues to selection, while ``with constrains" set $k = 90\% \cdot |\mathcal{F}|$.
}
\footnotesize
\label{tab:Preliminaryexperiment1}
\resizebox{0.85\linewidth}{!}{
\begin{tblr}{
  width = 0.9\linewidth,
  colspec = {Q[65]Q[146]Q[129]Q[146]Q[140]Q[129]Q[148]},
  cells = {c},
  cell{1}{1} = {r=2}{c},
  cell{1}{2} = {c=3}{0.4\linewidth},
  cell{1}{5} = {c=3}{0.4\linewidth},
  vline{1,2,5,8} = {-}{0.15em},
  vline{3-8} = {2-2}{0.15em},
  hline{1,3,10} = {-}{0.15em},
  hline{2} = {2-8}{0.15em},
}
scale $|\mathcal{F}|$ & without constraints &         &          & constraints ($k = 90\% \cdot |\mathcal{F}|$)  &           &          \\
      & SA                  & QA      & Hybrid   & SA                   & QA        & Hybrid   \\
10    & -0.053             & -0.052  & -0.053  & -0.051               & -0.051   & -0.051   \\
30    & -0.521             & -0.521 & -0.521  & -0.511                & -0.485    & -0.511   \\
50    & -1.834             & -1.834  & -1.835  & -1.758                & 2.3610    & -1.767   \\
100   & -4.988             & -4.988 & -4.988  & -4.621                & 18.494    & -4.831   \\
150   & -8.908             & -8.908 & -8.908  & -7.970                & 164.041   & -8.206   \\
300   & -17.541            & -       & -17.541 & -12.459              & -         & -13.331  \\
500   & -34.522            & -       & -34.523 & -17.214              & -         & -20.255 
\end{tblr}
}
\end{table}

\begin{figure}[htbp]
    \centering

    \begin{subfigure}[t]{0.47\textwidth}
        \centering
        \includegraphics[width=\textwidth]{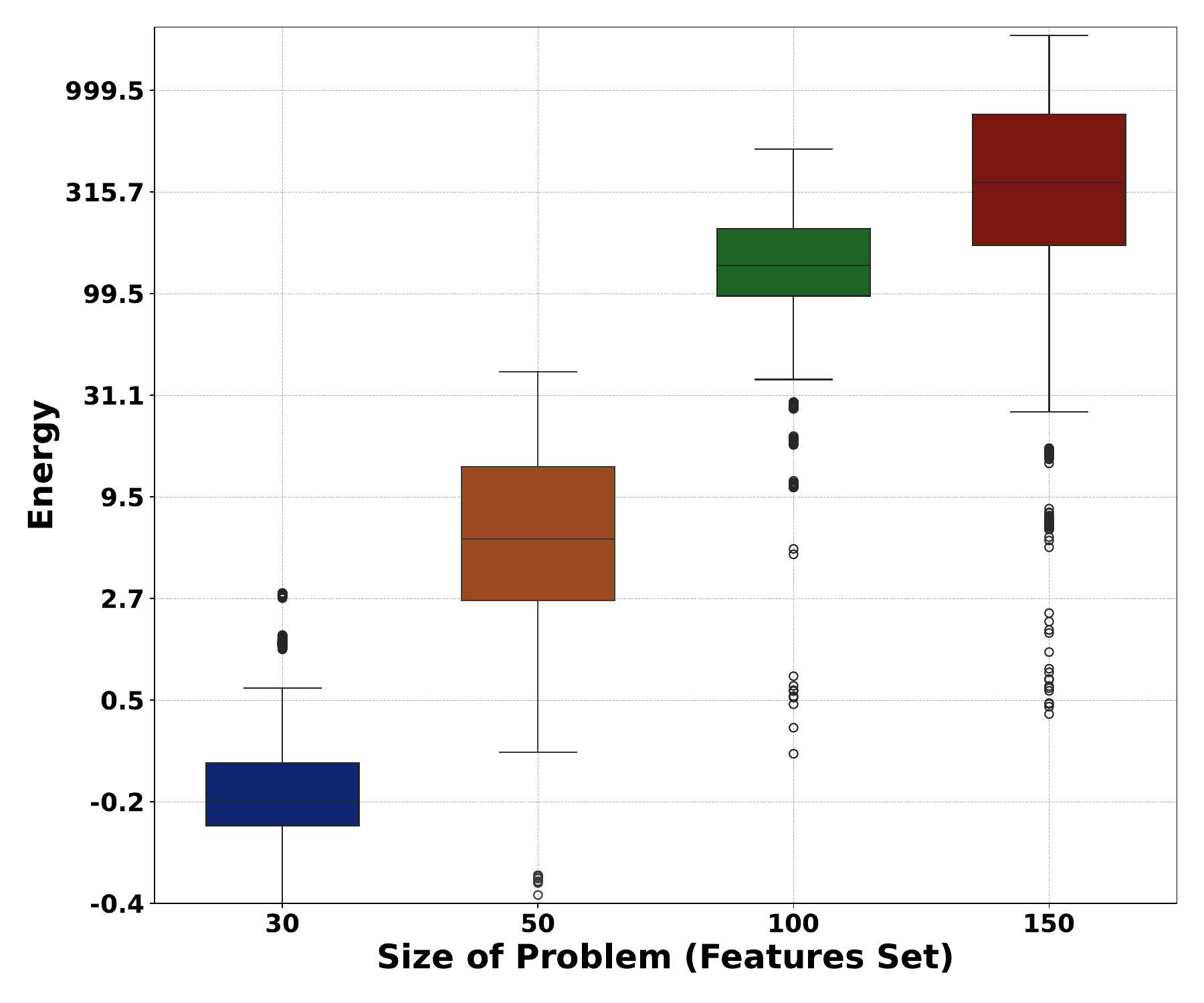}
        \caption{ Energy $Y$ vs problem size $|\mathcal{F}|$}
        \label{fig:sub3}
    \end{subfigure}
    \hspace{0.01\textwidth}
    \begin{subfigure}[t]{0.47\textwidth}
        \centering
        \includegraphics[width=\textwidth]{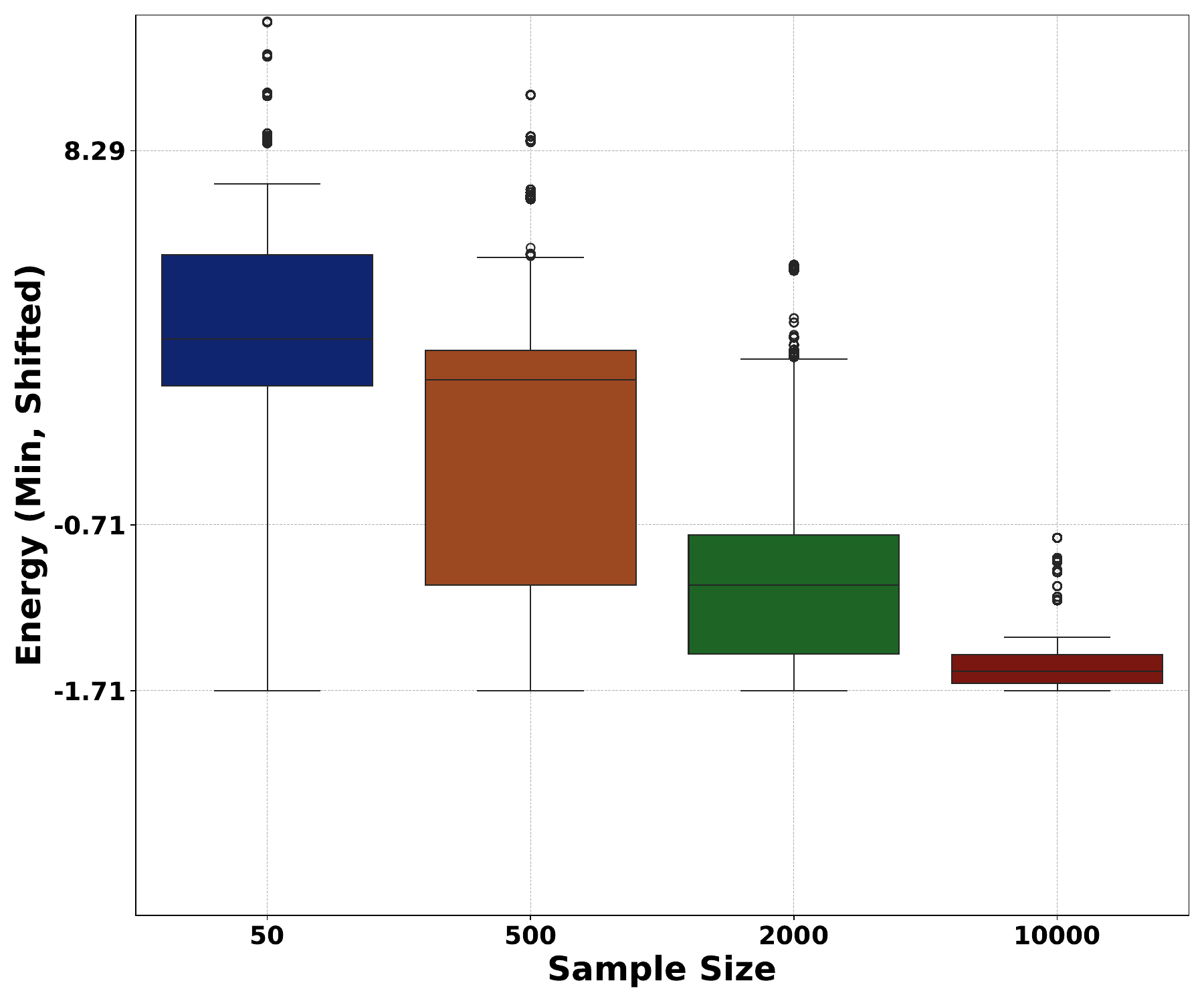}
        \caption{ Energy $Y$ vs sample size}
        \label{fig:sub4}
    \end{subfigure}
    
    \begin{subfigure}[t]{0.47\textwidth}
        \centering
        \includegraphics[width=\textwidth]{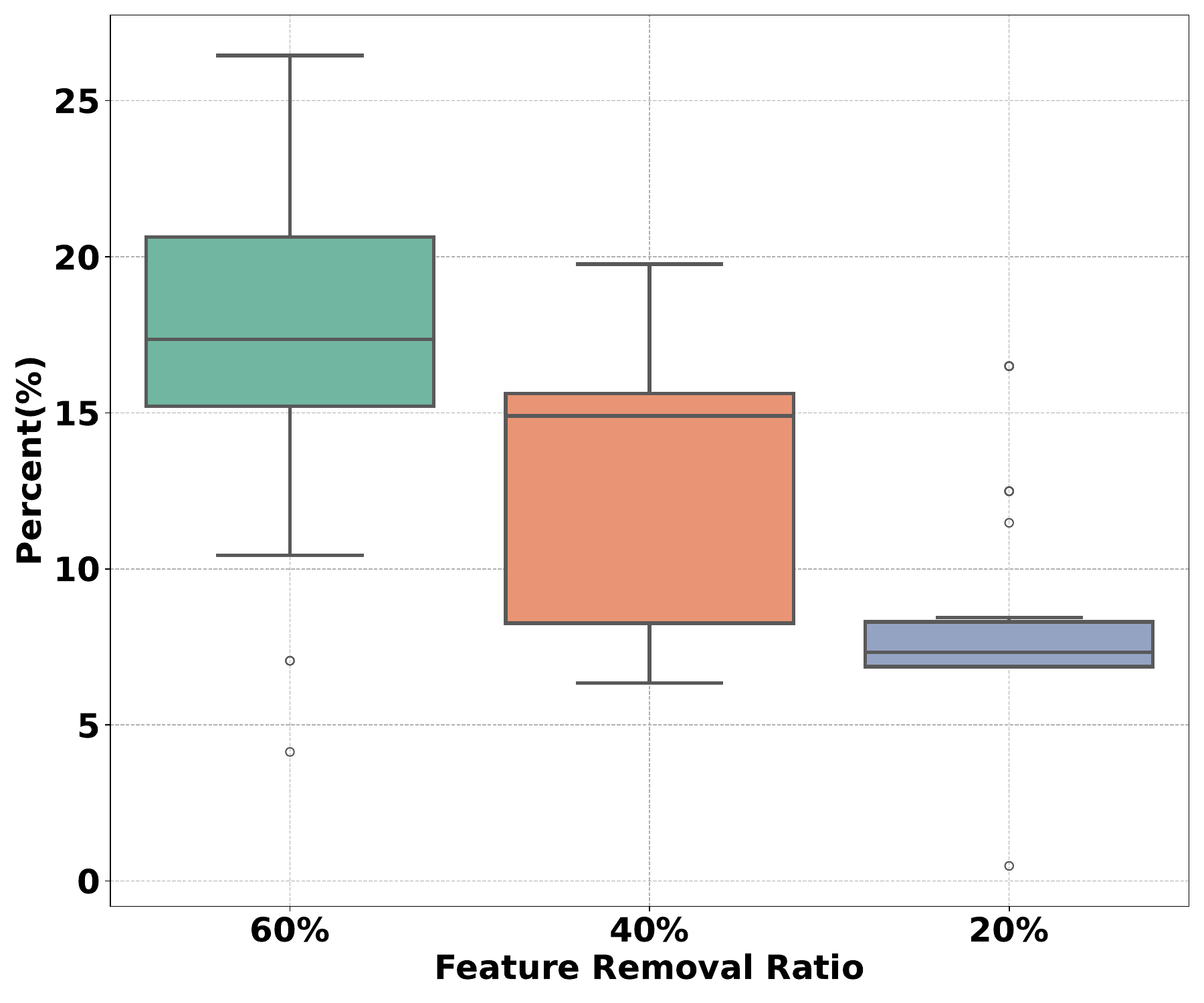}
        \caption{ Performance vs problem difficulty}
        \label{fig:diff-sparsity}
    \end{subfigure}
    \hspace{0.01\textwidth}
    \begin{subfigure}[t]{0.48\textwidth}
        \centering
        \includegraphics[width=\textwidth]{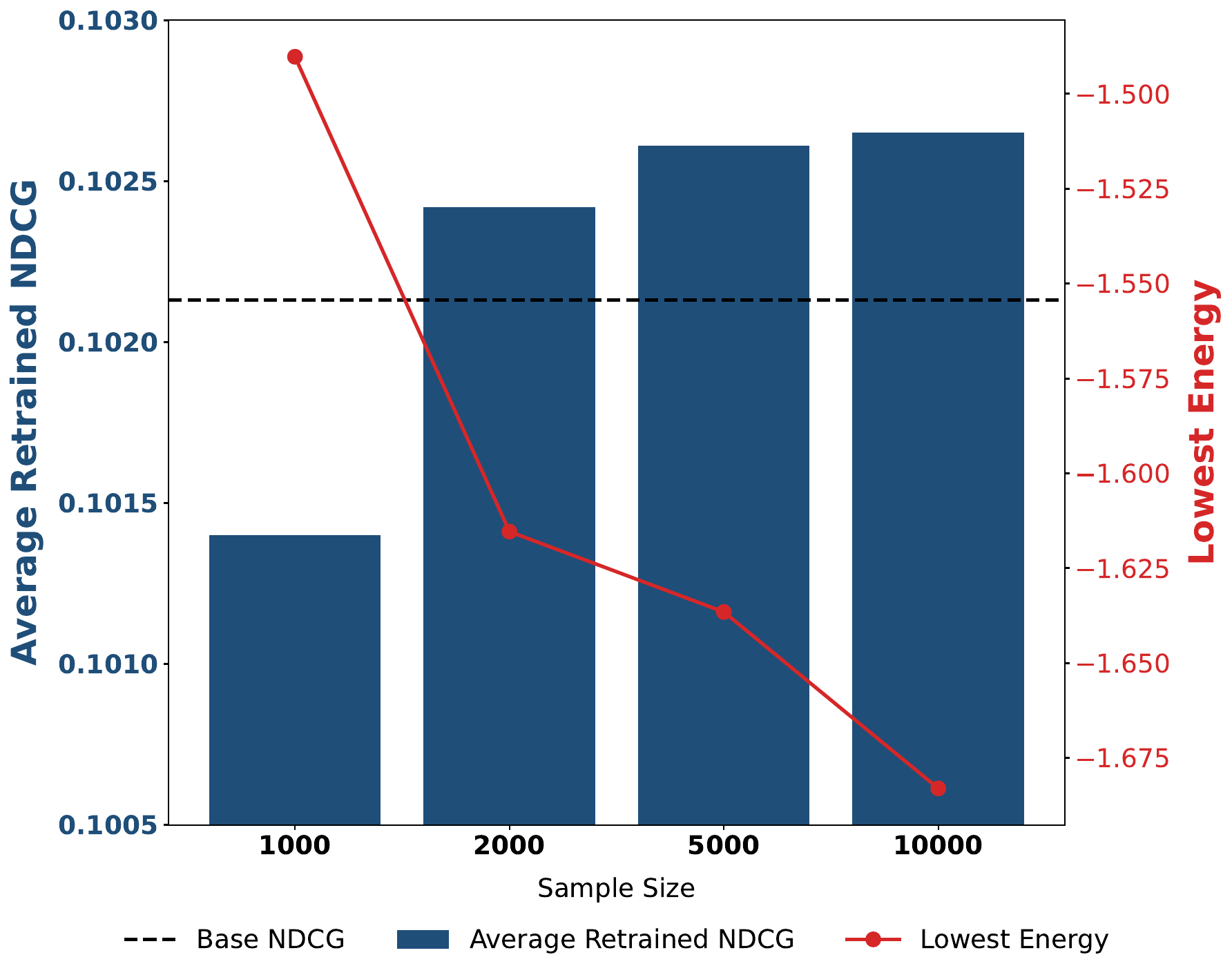}
        \caption{Performance vs sample size}
        \label{fig:performance-sample}
    \end{subfigure}

    \caption{
        (a) Distribution of energy values after selecting 90\% of the features using QA. 
        (b) Lowest energy values for a problem scale of 50, with 1000 repeated runs per sample size.
        (c) Recommendation performance vs problem difficulty, defined as the percentage of dropped feature values.
        (d) Performance vs sample size, using varying numbers of QA samples. Each point represents the lowest energy and retrained model performance, averaged over 10 runs.}
    \label{fig:second_group_2x1}
\end{figure}

\subsubsection{Performance Analysis}

\begin{figure}[htbp]
    \centering
    \begin{subfigure}[b]{0.47\textwidth} 
        \centering
        \includegraphics[width=\textwidth]{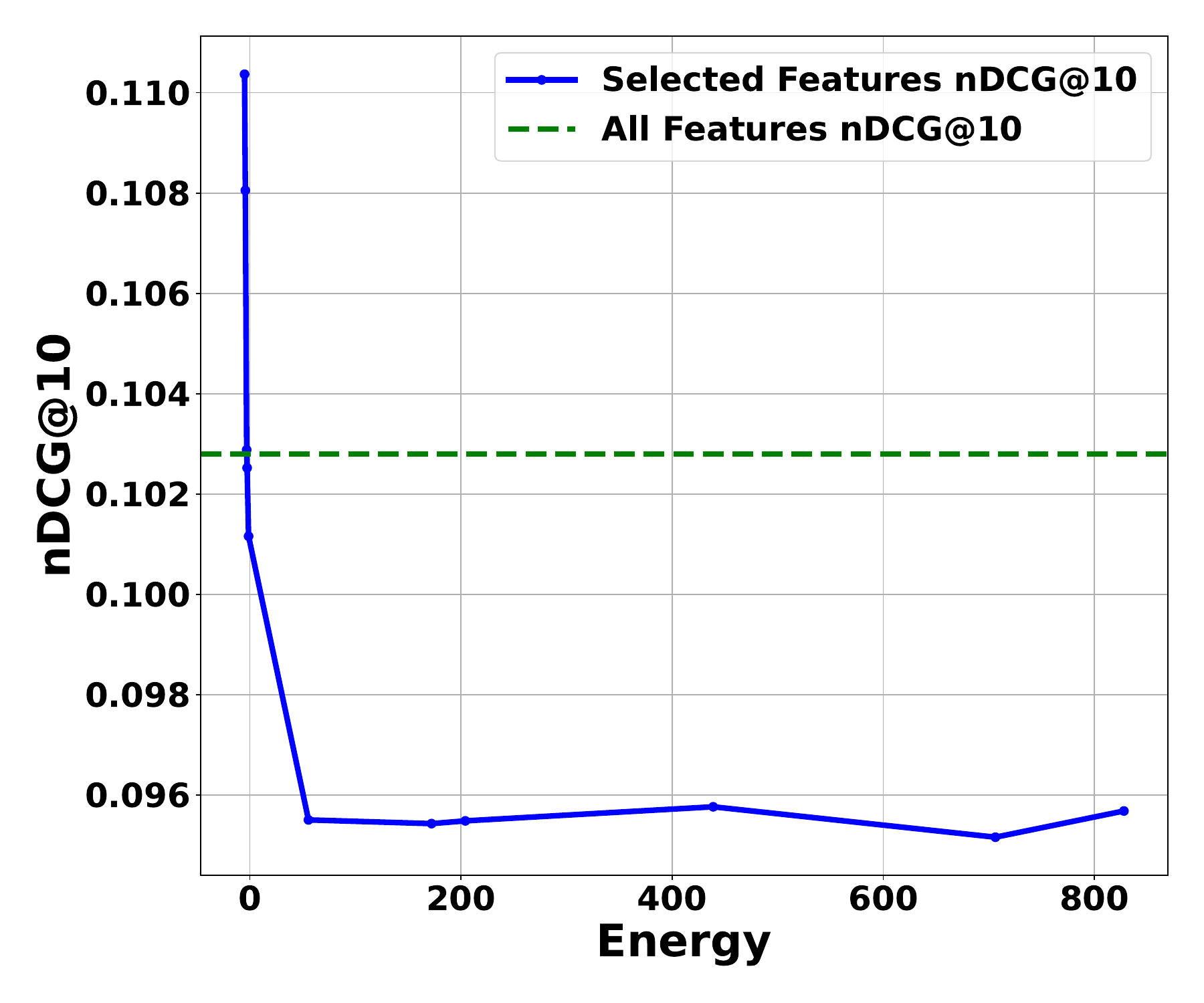}
        \caption{PDQUBO}
        \label{fig:sub1}
    \end{subfigure}
    \begin{subfigure}[b]{0.47\textwidth}
        \centering
        \includegraphics[width=\textwidth]{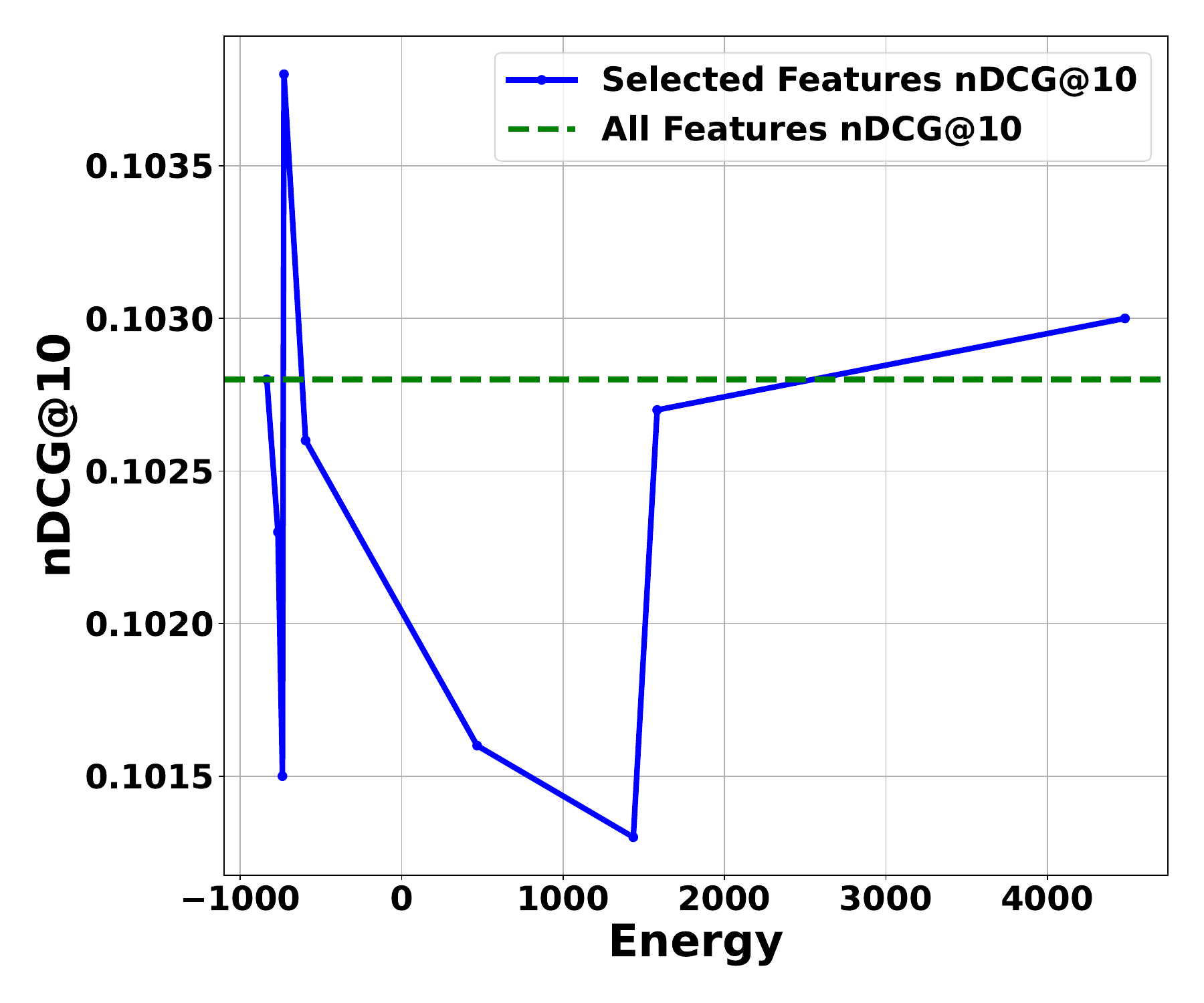}
        \caption{MIQUBO}
        \label{fig:sub2}
    \end{subfigure}
    \caption{Energy $Y$ vs nDCG@10}
    \label{fig:perform-analysis}
\end{figure}

In this section, we attempt to answer why and how the proposed PDQUBO drives the optimization direction of the QUBO problem toward the optimization of recommendation performance. First, the aim of QUBO is to minimize the energy value $Y$ as defined in Equation~\ref{eq:Equation2}. So, it is reasonable to assume: that because PDQUBO is performance-driven, there must be a clear correlation between the minimized $Y$ values and the final recommendation performance when using the corresponding selected feature set. In order to investigate this, we conducted an analytic experiment: 50 features were randomly selected from the 150ICM dataset, and PDQUBO and MIQUBO were deployed to select 45 features. 
The results are shown in Figure~\ref{fig:perform-analysis}. It is observed that there is a generally decreasing trend between energy and performance between $Y$ and the final performance measured in nDCG@10 from PDQUBO, while there is an unclear trend in that of MIQUBO. 
This demonstrates that considering the relationship between features and the ground truth (e.g. Mutual Information) in MIQUBO cannot perform as well as considering the relationship between features and the recommendation performance in PDQUBO.

\subsubsection{Stability of QA:}
\label{sec:Stability}
We conducted experiments to examine the current stability of the used QA. As shown in Table~\ref{tab:Preliminaryexperiment1}, where letting the optimization methods (QA, Hybrid, and SA) to decide how many features to select by themselves, their final optimization results are comparable. However, when giving a constrain by setting $k= 90\% * |\mathcal{F}|$ (namely selecting 90\% of features from $\mathcal{F}$), the corresponding $Y$ values of QA increases significantly when the scale (the size of the feature set $|\mathcal{F}|$) increases, which is different from other optimization methods, SA and Hybrid. We refers this as QA's instability, and we study this further from the following three perspectives: 
\begin{itemize}
    \item the size of the problem (the size of feature set $|\mathcal{F}|$): in D-wave, QA results are obtained by sampling 2,000 times and selecting the minimum $Y$. Figure~\ref{fig:sub3} shows how the distribution of $Y$ from single-sample QA varies with $|\mathcal{F}|$, indicating that current QA methods remain stable only for small problem sizes but struggle to find optimal solutions as the problem size increases.
    \item the size of the samples: we examine the effect of varying QA sample sizes on optimization stability and downstream performance. As shown in Figure~\ref{fig:sub4} and \ref{fig:performance-sample}, increasing the number of samples consistently reduces energy variance and helps the quantum annealer more reliably locate low-energy solutions. This improved stability leads to better feature subsets, which in turn enhance the recommendation model’s retrained performance.
    \item the difficulty of the problem: following~\cite{1423975}, we use data sparsity to measure the difficulty of the problem. Specifically, we set the size of the feature set to 50, then we randomly drop the feature values from all features with certain percentage to get various feature sets with various sparsities. Specifically, we randomly drop 60\%, 40\% and 20\% of feature values, then measure their corresponding relevant improvements to their corresponding recommendation performance with all feature performance (as shown in Figure~\ref{fig:diff-sparsity}). It is observed that QA tends to get larger but less stable improvement when the problem is harder (the feature set is sparser).
\end{itemize}

    

\begin{table}
\centering
\caption{Solving time vs. Problem scale. Note that the QA line is shorter due to quantum computational limitations in D-wave~\cite{McGeoch2020TheDA}.}
\label{tab:time}
\footnotesize
\resizebox{0.85\linewidth}{!}{
\begin{tblr}{
  width = \linewidth,
  colspec = {Q[137]Q[156]Q[156]Q[156]Q[131]Q[162]},
  row{even} = {c},
  row{3} = {c},
  row{5} = {c},
  row{7} = {c},
  row{9} = {c},
  cell{1}{1} = {r=2}{},
  cell{1}{2} = {c=3}{0.468\linewidth,c},
  cell{1}{5} = {c=2}{0.293\linewidth,c},
  vline{1,2,5,7} = {-}{0.15em},
  hline{1,3,10} = {-}{0.15em},  
}
\textbf{Scale } & \textbf{Tradition } &              &             & \textbf{QPU } &                 \\
                & \textbf{SA}         & \textbf{SGD} & \textbf{TS} & \textbf{QA}   & \textbf{Hybrid} \\
\textbf{10}     & 1.85                & 0.102        & 1.62        & 0.202         & 8.73            \\
\textbf{30}     & 2.26                & 0.832        & 2.17        & 0.178         & 9.6             \\
\textbf{50}     & 5.03                & 2.15         & 4.21        & 0.443         & 8.85            \\
\textbf{100}    & 13.65               & 8.55         & 12.51       & 0.496         & 9.51            \\
\textbf{150}    & 29.36               & 18.67        & 26.3        & 0.546         & 9.68            \\
\textbf{300}    & 138.93              & 75.48        & 133.58      & -             & 13.32           \\
\textbf{500}    & 391.03              & 212.19       & 472.01      & -             & 16.49           
\end{tblr}
}
\end{table}

\subsubsection{Efficiency of QUBO Optimization Methods:}
To highlight the speed advantage of quantum annealers, we compared the solving times of QA, Hybrid, and classical methods (such as SA, SGD, and TS) across different problem scales in Table~\ref{tab:time}. The recorded times for QA and Hybrid only include the quantum sampling time on the QPU and the post-processing time for decoding annealing results. For QA, the number of samples was set to 2000, meaning that the annealing duration for a single QA process was less than 1e-4 seconds, making the solving speed far superior to that of algorithms running on traditional computers. Furthermore, it can be observed that the solving times for the three classical methods increase exponentially with problem size, whereas the solving times for QA and Hybrid remain largely unaffected by problem scale.

\begin{table}[t]
\caption{``Indiv" vs ``all" in terms nDCG@10. ``Indiv" refers to using counterfactual instances that only remove one feature, while ``all" refers to the standard PDQUBO method.}
\centering
\footnotesize
\label{tab:Quadratic_Terms}
\resizebox{0.85\linewidth}{!}{
\begin{tblr}{
  width = \linewidth,
  colspec = {Q[121]Q[88]Q[63]Q[104]Q[106]Q[100]Q[102]},
  cells = {c},
  cell{1}{1} = {c=2,r=2}{0.209\linewidth},
  cell{1}{3} = {r=2}{},
  cell{1}{4} = {c=2}{0.21\linewidth},
  cell{1}{6} = {c=2}{0.202\linewidth},
  cell{3}{1} = {r=10}{},
  cell{3}{2} = {r=5}{},
  cell{8}{2} = {r=5}{},
  cell{13}{1} = {r=10}{},
  cell{13}{2} = {r=5}{},
  cell{18}{2} = {r=5}{},
  vline{1,8} = {-}{0.15em},
  vline{4} = {-}{0.1em},
  vline{2} = {2-23}{0.1em},
  vline{3} = {-}{0.1em},
  vline{6} = {-}{0.1em},
  hline{1,13,23} = {-}{0.15em},
  hline{2} = {4-8}{0.1em},
  hline{3} = {-}{0.1em},
  hline{8,18} = {2-8}{0.08em},
  hline{23} = {-}{0.15em},
}
Method    &        & $k$ & SA &                  &  Hybrid  &                 \\
          &        &       & Indiv               & all            & Indiv             & all            \\
KNN Model & 150ICM & \textasteriskcentered       & 0.1125              & \textbf{0.1168} & 0.1125            & \textbf{0.1168} \\
          &        & 130   & 0.1139              & \textbf{0.1140} & 0.1144            & \textbf{0.1151} \\
          &        & 135   & 0.1142              & \textbf{0.1156} & 0.1129            & \textbf{0.1137} \\
          &        & 140   & \textbf{0.1123}     & 0.1121 & 0.1128            & \textbf{0.1135} \\
          &        & 145   & 0.1115              & \textbf{0.1125} & 0.1115            & \textbf{0.1116} \\
          & 500ICM & \textasteriskcentered       & 0.1303              & \textbf{0.1359} & 0.1298            & \textbf{0.1359} \\
          &        & 350   & 0.1261              & \textbf{0.1279} & 0.1338            & \textbf{0.1339} \\
          &        & 400   & \textbf{0.1331}     & 0.1320          & 0.1336            & \textbf{0.1370} \\
          &        & 450   & 0.1332              & \textbf{0.1339} & 0.1368            & \textbf{0.1370} \\
          &        & 470   & 0.1321              & \textbf{0.1373} & 0.1359            & \textbf{0.1402} \\
MLP-DP Model & 150ICM & \textasteriskcentered       & 0.1240              & \textbf{0.1299} & 0.1240            & \textbf{0.1299} \\
          &        & 130   & 0.1332              & \textbf{0.1366} & 0.1316            & \textbf{0.1346} \\
          &        & 135   & 0.1335              & \textbf{0.1350} & 0.1337            & \textbf{0.1350} \\
          &        & 140   & 0.1334              & \textbf{0.1356} & 0.1351            & \textbf{0.1355} \\
          &        & 145   & 0.1321              & \textbf{0.1334} & 0.1321            & \textbf{0.1322} \\
          & 500ICM & \textasteriskcentered       & 0.1272              & \textbf{0.1365} & 0.1272            & \textbf{0.1365} \\
          &        & 350   & 0.1378              & \textbf{0.1393} & 0.1376            & \textbf{0.1391} \\
          &        & 400   & 0.1389              & \textbf{0.1430} & 0.1411            & \textbf{0.1441} \\
          &        & 450   & 0.1416              & \textbf{0.1424} & 0.1420            & \textbf{0.1424} \\
          &        & 470   & 0.1433              & \textbf{0.1449} & 0.1418            & \textbf{0.1428} 
\end{tblr}
}
\end{table}

\subsubsection{Necessity of Quadratic Terms:}\label{sec-exp-quadraticterm}

Comparing the effects of removing two features simultaneously (``comb") versus removing a single feature (``Indiv") is essential, as it helps determine whether the current problem can be formulated as a QUBO problem and if a quantum annealer is necessary for solving it. The inter dependencies between features suggest that removing two features at once produces more accurate counterfactual instances for feature selection than removing only one. We selected two base models (Item-KNN and MLP-DP) to evaluate the performance of feature selection by comparing the use of only the diagonal elements of matrix $Q$ (representing counterfactual instances generated by the absence of a single feature) against the use of the full matrix $Q$ (representing counterfactual instances generated by the simultaneous removal of two features). This comparison was performed using both SA and Hybrid optimization methods, and the results are shown in Table~\ref{tab:Quadratic_Terms}.
When $k=*$, the ``all" method outperformed the ``Indiv" method for both the Item-KNN and MLP-DP. However, in cases of $k \neq *$, there were a few cases where the ``Indiv" method produced better performance. This may be due to the fact that in the ``Indiv" approach, the off-diagonal elements of the $Q$ are zero, simplifying the optimization process. 

To further explore this, we conducted additional experiments using the Hybrid method, selecting 135 features for the Item-KNN with both ``all" and ``Indiv" approaches, repeating the process 150 times. We then calculated nDCG@10 for the Item-KNN model and performed a Kolmogorov–Smirnov (KS) test to determine if there were significant differences between the performance distributions. The results showed a mean nDCG@10 of 0.1151 for ``all" and 0.1139 for ``Indiv", with the KS statistic $= 0.2465$ and $p$-value $= 7.84\mathrm{e}{-5}$. These findings indicate a significant difference between the two approaches, with the ``all" method demonstrating superior performance over the ``Indiv" method.

\subsubsection{Fully Retrained Evaluation on Industrial Datasets:}
\label{sec:retrain_eval}

The fixed-parameter evaluation in Table~\ref{tab:table1} measures the intrinsic effect of feature selection under a shared trained model configuration.  However, practical recommender systems typically retrain models after feature selection and re-optimize hyperparameters. To assess PDQUBO under such deployment-oriented conditions, we conduct a fully retrained evaluation on 150\_ICM, 500\_ICM, KuaiRec, and KuaiRand. KuaiRec contains 116 item features (1,411 users and 3,276 items), split into 138,992/34,748/43,435 train/validation/test interactions. KuaiRand contains 224 item features (13,077 users and 3,359 items), with 257,540/58,264/59,911 interactions in the respective splits. These datasets represent industrial-scale recommendation scenarios with heterogeneous user–item distributions. 

In this setting, QUBO instances are solved using Simulated Annealing (SA). After feature selection, all base models are retrained from scratch and hyperparameters are re-optimized via Bayesian optimization (200 iterations, including 40 initialization trials). This protocol enables adaptive capacity adjustment for each selected feature subset and reflects realistic deployment procedures. Overall, Bayesian re-optimization consistently improves the ALL-feature baselines (e.g., MLP-DP increases from 0.1308 to 0.1422 on 150\_ICM and from 0.1401 to 0.1468 on 500\_ICM).  More importantly, the stability of feature selection improves markedly. Configurations that previously underperformed relative to the ALL-feature model (e.g., MLP-DP and NCF at $k=350$ on 500\_ICM, and MLP-CON at $k=130$–$140$ on 150\_ICM) are largely resolved after retraining. Only one setting (MLP-CON at $k=145$) remains below the ALL-feature baseline. Furthermore, the magnitude of improvement after feature selection becomes more pronounced. For instance, under the NCF model on 500\_ICM, performance increases from 0.1388 (ALL features) to 0.1491 after selection and retraining. Across datasets and base models, PDQUBO achieves the best or second-best performance in the majority of configurations, while no method universally dominates all tasks, consistent with observations in feature selection benchmarks.

These results indicate that PDQUBO remains effective under adaptive retraining and scales robustly to industrial datasets, demonstrating both intrinsic selection quality and practical deployment viability.

\begin{table}
\centering
\caption{
\label{tab:retrain_bayesian}
Fully retrained evaluation results optimized with Simulated Annealing (SA) for QUBO solving. After feature selection, all base models are retrained and hyperparameters are re-optimized via Bayesian optimization. The reported metric is nDCG@10. Results marked in \textbf{\color{red} red bold} indicate the best performance under each setting, while \underline{\textbf{black bold}} denotes the second-best performance. “ALL” represents the performance of the base model using all available features without feature selection. All results are averaged over five runs.
}
\resizebox{1.0\linewidth}{!}{
\tiny
\begin{tblr}{
  width = \linewidth,
  colspec = {Q[110]Q[50]Q[50]Q[50]Q[50]Q[50]Q[50]Q[50]Q[50]Q[55]Q[50]Q[50]Q[50]Q[70]Q[50]Q[50]Q[50]},
  cells = {c},
  cell{1}{1} = {r=2}{},
  cell{3}{2} = {r=5}{},
  cell{3}{4} = {font=\bfseries},
  cell{3}{6} = {r=5}{},
  cell{3}{7} = {font=\bfseries},
  cell{3}{10} = {r=5}{},
  cell{3}{14} = {r=5}{},
  cell{3}{17} = {font=\bfseries},
  cell{4}{11} = {font=\bfseries},
  cell{4}{15} = {font=\bfseries},
  cell{5}{3} = {fg=red,font=\bfseries},
  cell{5}{5} = {font=\bfseries},
  cell{5}{7} = {fg=red,font=\bfseries},
  cell{5}{8} = {fg=red,font=\bfseries},
  cell{5}{9} = {font=\bfseries},
  cell{5}{13} = {font=\bfseries},
  cell{6}{12} = {fg=red,font=\bfseries},
  cell{6}{16} = {fg=red,font=\bfseries},
  cell{7}{3} = {font=\bfseries},
  cell{7}{4} = {fg=red,font=\bfseries},
  cell{7}{5} = {fg=red,font=\bfseries},
  cell{7}{8} = {font=\bfseries},
  cell{7}{9} = {fg=red,font=\bfseries},
  cell{7}{11} = {fg=red,font=\bfseries},
  cell{7}{12} = {font=\bfseries},
  cell{7}{13} = {fg=red,font=\bfseries},
  cell{7}{15} = {fg=red,font=\bfseries},
  cell{7}{16} = {font=\bfseries},
  cell{7}{17} = {fg=red,font=\bfseries},
  cell{8}{2} = {r=5}{},
  cell{8}{3} = {font=\bfseries},
  cell{8}{4} = {fg=red,font=\bfseries},
  cell{8}{6} = {r=5}{},
  cell{8}{10} = {r=5}{},
  cell{8}{11} = {fg=red,font=\bfseries},
  cell{8}{13} = {fg=red,font=\bfseries},
  cell{8}{14} = {r=5}{},
  cell{8}{15} = {font=\bfseries},
  cell{8}{16} = {font=\bfseries},
  cell{9}{4} = {font=\bfseries},
  cell{9}{5} = {font=\bfseries},
  cell{9}{7} = {fg=red,font=\bfseries},
  cell{9}{8} = {font=\bfseries},
  cell{9}{12} = {font=\bfseries},
  cell{10}{9} = {font=\bfseries},
  cell{11}{9} = {fg=red,font=\bfseries},
  cell{11}{17} = {fg=red,font=\bfseries},
  cell{12}{3} = {fg=red,font=\bfseries},
  cell{12}{5} = {fg=red,font=\bfseries},
  cell{12}{7} = {font=\bfseries},
  cell{12}{8} = {fg=red,font=\bfseries},
  cell{12}{11} = {font=\bfseries},
  cell{12}{12} = {fg=red,font=\bfseries},
  cell{12}{13} = {font=\bfseries},
  cell{12}{15} = {fg=red,font=\bfseries},
  cell{12}{16} = {fg=red,font=\bfseries},
  cell{12}{17} = {font=\bfseries},
  cell{13}{2} = {r=5}{},
  cell{13}{3} = {font=\bfseries},
  cell{13}{4} = {font=\bfseries},
  cell{13}{6} = {r=5}{},
  cell{13}{8} = {font=\bfseries},
  cell{13}{10} = {r=5}{},
  cell{13}{14} = {r=5}{},
  cell{14}{13} = {font=\bfseries},
  cell{15}{5} = {font=\bfseries},
  cell{15}{11} = {font=\bfseries},
  cell{16}{7} = {fg=red,font=\bfseries},
  cell{16}{9} = {font=\bfseries},
  cell{16}{12} = {font=\bfseries},
  cell{16}{16} = {font=\bfseries},
  cell{16}{17} = {fg=red,font=\bfseries},
  cell{17}{3} = {fg=red,font=\bfseries},
  cell{17}{4} = {fg=red,font=\bfseries},
  cell{17}{5} = {fg=red,font=\bfseries},
  cell{17}{7} = {font=\bfseries},
  cell{17}{8} = {fg=red,font=\bfseries},
  cell{17}{9} = {fg=red,font=\bfseries},
  cell{17}{11} = {fg=red,font=\bfseries},
  cell{17}{12} = {fg=red,font=\bfseries},
  cell{17}{13} = {fg=red,font=\bfseries},
  cell{17}{15} = {font=\bfseries},
  cell{17}{16} = {fg=red,font=\bfseries},
  cell{17}{17} = {font=\bfseries},
  cell{18}{2} = {r=5}{},
  cell{18}{3} = {font=\bfseries},
  cell{18}{4} = {fg=red,font=\bfseries},
  cell{18}{6} = {r=5}{},
  cell{18}{10} = {r=5}{},
  cell{18}{11} = {fg=red,font=\bfseries},
  cell{18}{12} = {font=\bfseries},
  cell{18}{14} = {r=5}{},
  cell{18}{15} = {fg=red,font=\bfseries},
  cell{18}{17} = {font=\bfseries},
  cell{19}{7} = {font=\bfseries},
  cell{19}{8} = {font=\bfseries},
  cell{19}{9} = {font=\bfseries},
  cell{20}{13} = {font=\bfseries},
  cell{21}{16} = {fg=red,font=\bfseries},
  cell{22}{3} = {fg=red,font=\bfseries},
  cell{22}{4} = {font=\bfseries},
  cell{22}{7} = {fg=red,font=\bfseries},
  cell{22}{8} = {fg=red,font=\bfseries},
  cell{22}{9} = {fg=red,font=\bfseries},
  cell{22}{11} = {font=\bfseries},
  cell{22}{12} = {fg=red,font=\bfseries},
  cell{22}{13} = {fg=red,font=\bfseries},
  cell{22}{15} = {fg=red,font=\bfseries},
  cell{22}{16} = {fg=red,font=\bfseries},
  cell{22}{17} = {fg=red,font=\bfseries},
  hline{1, 3, 23} = {-}{0.15em},
  hline{8, 13, 18} = {-}{0.08em},
  hline{2} = {2-18}{0.08em},
  vline{1, 18} = {-}{0.15em},
  vline{2, 3, 6, 7, 10, 11, 14, 15} = {-}{0.08em},
}
Method        & ICM150 & MLP-DP         & NCF            & MLP-CON            & ICM500 & MLP-DP         & NCF            & MLP-CON           & KuaiRec  & MLP-DP         & NCF            & MLP-CON            & KuaiRand & MLP-DP         & NCF            & MLP-CON            \\
              & ALL    & 0.1422         & 0.1346         & 0.1184         & ALL~   & 0.1468         & 0.1388         & 0.1105         & ALL     & 0.2473         & 0.2542         & 0.2327         & ALL      & 0.0350         & 0.0343         & 0.0383         \\
CQFS          & 130    & 0.1323         & \uline{0.1382} & 0.1153         & 350    & \uline{0.1486} & 0.1424         & 0.1105         & 92      & 0.2475         & 0.2478         & 0.2297         & 179      & 0.0325         & 0.0350         & \uline{0.0393} \\
QUBO-BST      &        & 0.1414         & 0.1373         & 0.1170         &        & 0.1448         & 0.1397         & 0.1110         &         & \uline{0.2481} & 0.2437         & 0.2299         &          & \uline{0.0354} & 0.0350         & 0.0378         \\
CoQUBO        &        & 0.1455         & 0.1363         & \uline{0.1185} &        & 0.1493         & 0.1481         & \uline{0.1145} &         & 0.2455         & 0.2419         & \uline{0.2316} &          & 0.0353         & 0.0339         & 0.0387         \\
MIQUBO        &        & 0.1414         & 0.1330         & 0.1158         &        & 0.1449         & 0.1375         & 0.1131         &         & 0.2153         & 0.2503         & 0.2279         &          & 0.0326         & 0.0360         & 0.0392         \\
PBQUBO        &        & \uline{0.1424} & 0.1420         & 0.1207         &        & 0.1465         & \uline{0.1472} & 0.1166         &         & 0.2502         & \uline{0.2489} & 0.2375         &          & 0.0355         & \uline{0.0355} & 0.0396         \\
CQFS          & 135    & \uline{0.1473} & 0.1399         & 0.1192         & 400    & 0.1379         & 0.1363         & 0.1100         & 98      & 0.2503         & 0.2441         & 0.2370         & 190      & \uline{0.0355} & \uline{0.0343} & 0.0369         \\
QUBO-BST      &        & 0.1408         & \uline{0.1386} & \uline{0.1196} &        & 0.1487         & \uline{0.1468} & 0.1108         &         & 0.2451         & \uline{0.2472} & 0.2332         &          & 0.0349         & 0.0331         & 0.0377         \\
CoQUBO        &        & 0.1442         & 0.1369         & 0.1120         &        & 0.1420         & 0.1408         & \uline{0.1146} &         & 0.2458         & 0.2467         & 0.2330         &          & 0.0341         & 0.0332         & 0.0378         \\
MIQUBO        &        & 0.1446         & 0.1356         & 0.1145         &        & 0.1452         & 0.1389         & 0.1150         &         & 0.2264         & 0.2417         & 0.2192         &          & 0.0351         & 0.0341         & 0.0397         \\
PBQUBO        &        & 0.1474         & 0.1343         & 0.1207         &        & \uline{0.1473} & 0.1491         & 0.1107         &         & \uline{0.2495} & 0.2552         & \uline{0.2349} &          & 0.0368         & 0.0352         & \uline{0.0394} \\
CQFS          & 140    & \uline{0.1442} & \uline{0.1407} & 0.1106         & 450    & 0.1458         & \uline{0.1438} & 0.1134         & 104     & 0.2458         & 0.2507         & 0.2331         & 201      & 0.0333         & 0.0321         & 0.0373         \\
QUBO-BST      &        & 0.1398         & 0.1298         & 0.1171         &        & 0.1411         & 0.1437         & 0.1102         &         & 0.2458         & 0.2507         & \uline{0.2332} &          & 0.0368         & 0.0334         & 0.0374         \\
CoQUBO        &        & 0.1410         & 0.1392         & \uline{0.1205} &        & 0.1438         & 0.1437         & 0.1107         &         & \uline{0.2466} & 0.2525         & 0.2302         &          & 0.0351         & 0.0335         & 0.0376         \\
MIQUBO        &        & 0.1396         & 0.1396         & 0.1120         &        & 0.1507         & 0.1383         & \uline{0.1179} &         & 0.2264         & \uline{0.2535} & 0.2293         &          & 0.0329         & \uline{0.0349} & 0.0393         \\
PBQUBO        &        & 0.1474         & 0.1426         & 0.1232         &        & \uline{0.1476} & 0.1438         & 0.1232         &         & 0.2478         & 0.2552         & 0.2365         &          & \uline{0.0365} & 0.0353         & \uline{0.0390} \\
CQFS          & 145    & \uline{0.1502} & 0.1435         & 0.1107         & 470    & 0.1464         & 0.1388         & 0.1110         & 110     & 0.2489         & \uline{0.2528} & 0.2310         & 212      & 0.0359         & 0.0337         & \uline{0.0382} \\
QUBO-BST      &        & 0.1391         & 0.1346         & 0.1153         &        & \uline{0.1475} & \uline{0.1422} & \uline{0.1136} &         & 0.2474         & 0.2476         & 0.2322         &          & 0.0356         & 0.0336         & 0.0374         \\
CoQUBO        &        & 0.1391         & 0.1382         & 0.1117         &        & 0.1398         & 0.1399         & 0.1116         &         & 0.2464         & 0.2517         & \uline{0.2331} &          & 0.0354         & 0.0338         & 0.0374         \\
MIQUBO        &        & 0.1436         & 0.1370         & 0.1132         &        & 0.1465         & 0.1420         & 0.1109         &         & 0.2288         & 0.2472         & 0.2294         &          & 0.0351         & 0.0351         & 0.0379         \\
PBQUBO        &        & 0.1513         & \uline{0.1399} & 0.1177         &        & 0.1499         & 0.1430         & 0.1177         &         & \uline{0.2484} & 0.2555         & 0.2386         &          & 0.0359         & 0.0351         & 0.0391      
\end{tblr}
}
\end{table}

\subsubsection{Scalability Analysis and Two-Stage Acceleration:} The main computational cost of PDQUBO lies in constructing the QUBO coefficient matrix. For a feature set $\mathcal{F}$, the full formulation requires
\[
|\mathcal{F}| + \frac{|\mathcal{F}|(|\mathcal{F}|-1)}{2}
\]
counterfactual evaluations of the trained base model, resulting in $O(|\mathcal{F}|^2)$ complexity. This cost arises from explicitly modeling pairwise feature interactions.

Since QUBO construction is performed after training, all counterfactual evaluations correspond to independent forward inference passes. Therefore, the most direct acceleration strategy is parallelization: these evaluations can be distributed across multiple GPUs without affecting correctness. Beyond hardware parallelism, we investigate a two-stage pruning strategy to reduce quadratic evaluations. In Stage 1, features are ranked by their individual counterfactual contributions (diagonal terms), and a subset is pre-selected for removal. In Stage 2, pairwise interactions are constructed only among the remaining features, and the reduced QUBO is optimized.

To measure how closely this strategy approximates full PDQUBO, we define
\[
\text{Remove-Jaccard}
=
\frac{|R_{\text{two-stage}} \cap R_{\text{full}}|}
{|R_{\text{two-stage}} \cup R_{\text{full}}|},
\]
where $R_{\text{two-stage}}$ and $R_{\text{full}}$ denote the removed feature sets of the two-stage and full PDQUBO, respectively. We additionally include a control variant (PDQUBO Random), where Stage 1 selects features randomly. Experiments are conducted on ICM500 and Ku ai R an d, the datasets with the largest feature dimensions in this study, under the same settings as Sec.~\ref{sec:retrain_eval}. We report post-selection nDCG and Remove-Jaccard in Table~\ref{tab:Two_stage}. The results show that full PDQUBO consistently achieves the best performance. Both the two-stage and random-first-stage variants underperform the full formulation, and their Remove-Jaccard values are generally low. Moreover, the two-stage strategy does not significantly outperform the random baseline, indicating that individual counterfactual contributions alone are insufficient to approximate the interaction-aware solution. These findings suggest that the quadratic interaction modeling is essential to PDQUBO. While two-stage pruning reduces computation, it does not reliably preserve the combinatorial structure captured by the full formulation.

\begin{table}
\centering
\caption{Comparison of full PDQUBO and two-stage pruning strategies on ICM500 and KuaiRand. We report post-selection nDCG and Remove-Jaccard (overlap with full PDQUBO removed features). “Random” and “Two-Stage” denote different first-stage pruning strategies. Red and underlined values represent the best and second-best results, respectively.}
\label{tab:Two_stage}
\resizebox{0.85\linewidth}{!}{
\footnotesize
\begin{tblr}{
  width = \linewidth,
  colspec = {Q[204]Q[119]Q[90]Q[77]Q[77]Q[117]Q[90]Q[77]Q[77]},
  cells = {c},
  cell{1}{1} = {r=2}{},
  cell{3}{2} = {r=5}{},
  cell{3}{4} = {font=\bfseries},
  cell{3}{5} = {font=\bfseries},
  cell{3}{6} = {r=5}{},
  cell{3}{7} = {font=\bfseries},
  cell{3}{8} = {font=\bfseries},
  cell{4}{3} = {fg=blue,font=\bfseries\itshape},
  cell{4}{4} = {fg=blue,font=\bfseries\itshape},
  cell{4}{5} = {fg=blue,font=\bfseries\itshape},
  cell{4}{7} = {fg=blue,font=\bfseries\itshape},
  cell{4}{8} = {fg=blue,font=\bfseries\itshape},
  cell{4}{9} = {fg=blue,font=\bfseries\itshape},
  cell{5}{8} = {font=\bfseries},
  cell{5}{9} = {font=\bfseries},
  cell{6}{3} = {fg=blue,font=\bfseries\itshape},
  cell{6}{4} = {fg=blue,font=\bfseries\itshape},
  cell{6}{5} = {fg=blue,font=\bfseries\itshape},
  cell{6}{7} = {fg=blue,font=\bfseries\itshape},
  cell{6}{8} = {fg=blue,font=\bfseries\itshape},
  cell{6}{9} = {fg=blue,font=\bfseries\itshape},
  cell{7}{4} = {fg=red,font=\bfseries},
  cell{7}{5} = {fg=red,font=\bfseries},
  cell{7}{7} = {fg=red,font=\bfseries},
  cell{7}{8} = {fg=red,font=\bfseries},
  cell{7}{9} = {fg=red,font=\bfseries},
  cell{8}{2} = {r=5}{},
  cell{8}{3} = {fg=red,font=\bfseries},
  cell{8}{4} = {font=\bfseries},
  cell{8}{5} = {fg=red,font=\bfseries},
  cell{8}{6} = {r=5}{},
  cell{8}{7} = {font=\bfseries},
  cell{8}{8} = {font=\bfseries},
  cell{8}{9} = {font=\bfseries},
  cell{9}{3} = {fg=blue,font=\bfseries\itshape},
  cell{9}{4} = {fg=blue,font=\bfseries\itshape},
  cell{9}{5} = {fg=blue,font=\bfseries\itshape},
  cell{9}{7} = {fg=blue,font=\bfseries\itshape},
  cell{9}{8} = {fg=blue,font=\bfseries\itshape},
  cell{9}{9} = {fg=blue,font=\bfseries\itshape},
  cell{11}{3} = {fg=blue,font=\bfseries\itshape},
  cell{11}{4} = {fg=blue,font=\bfseries\itshape},
  cell{11}{5} = {fg=blue,font=\bfseries\itshape},
  cell{11}{7} = {fg=blue,font=\bfseries\itshape},
  cell{11}{8} = {fg=blue,font=\bfseries\itshape},
  cell{11}{9} = {fg=blue,font=\bfseries\itshape},
  cell{12}{3} = {font=\bfseries},
  cell{12}{4} = {fg=red,font=\bfseries},
  cell{12}{5} = {font=\bfseries},
  cell{12}{7} = {fg=red,font=\bfseries},
  cell{12}{8} = {fg=red,font=\bfseries},
  cell{12}{9} = {fg=red,font=\bfseries},
  cell{13}{2} = {r=5}{},
  cell{13}{5} = {font=\bfseries},
  cell{13}{6} = {r=5}{},
  cell{14}{3} = {fg=blue,font=\bfseries\itshape},
  cell{14}{4} = {fg=blue,font=\bfseries\itshape},
  cell{14}{5} = {fg=blue,font=\bfseries\itshape},
  cell{14}{7} = {fg=blue,font=\bfseries\itshape},
  cell{14}{8} = {fg=blue,font=\bfseries\itshape},
  cell{14}{9} = {fg=blue,font=\bfseries\itshape},
  cell{15}{3} = {font=\bfseries},
  cell{15}{4} = {font=\bfseries},
  cell{15}{7} = {font=\bfseries},
  cell{15}{8} = {font=\bfseries},
  cell{15}{9} = {font=\bfseries},
  cell{16}{3} = {fg=blue,font=\bfseries\itshape},
  cell{16}{4} = {fg=blue,font=\bfseries\itshape},
  cell{16}{5} = {fg=blue,font=\bfseries\itshape},
  cell{16}{7} = {fg=blue,font=\bfseries\itshape},
  cell{16}{8} = {fg=blue,font=\bfseries\itshape},
  cell{16}{9} = {fg=blue,font=\bfseries\itshape},
  cell{17}{3} = {fg=red,font=\bfseries},
  cell{17}{4} = {fg=red,font=\bfseries},
  cell{17}{5} = {fg=red,font=\bfseries},
  cell{17}{7} = {fg=red,font=\bfseries},
  cell{17}{8} = {fg=red,font=\bfseries},
  cell{17}{9} = {fg=red,font=\bfseries},
  cell{18}{2} = {r=5}{},
  cell{18}{6} = {r=5}{},
  cell{18}{8} = {font=\bfseries},
  cell{19}{3} = {fg=blue,font=\bfseries\itshape},
  cell{19}{4} = {fg=blue,font=\bfseries\itshape},
  cell{19}{5} = {fg=blue,font=\bfseries\itshape},
  cell{19}{7} = {fg=blue,font=\bfseries\itshape},
  cell{19}{8} = {fg=blue,font=\bfseries\itshape},
  cell{19}{9} = {fg=blue,font=\bfseries\itshape},
  cell{20}{3} = {font=\bfseries},
  cell{20}{4} = {font=\bfseries},
  cell{20}{5} = {font=\bfseries},
  cell{20}{7} = {font=\bfseries},
  cell{20}{9} = {font=\bfseries},
  cell{21}{3} = {fg=blue,font=\bfseries\itshape},
  cell{21}{4} = {fg=blue,font=\bfseries\itshape},
  cell{21}{5} = {fg=blue,font=\bfseries\itshape},
  cell{21}{7} = {fg=blue,font=\bfseries\itshape},
  cell{21}{8} = {fg=blue,font=\bfseries\itshape},
  cell{21}{9} = {fg=blue,font=\bfseries\itshape},
  cell{22}{3} = {fg=red,font=\bfseries},
  cell{22}{4} = {fg=red,font=\bfseries},
  cell{22}{5} = {fg=red,font=\bfseries},
  cell{22}{7} = {fg=red,font=\bfseries},
  cell{22}{8} = {fg=red,font=\bfseries},
  cell{22}{9} = {fg=red,font=\bfseries},
  hline{1, 3, 23} = {-}{0.15em},
  hline{2} = {2-10}{0.08em},
  hline{8, 13, 18} = {-}{0.08em},
  vline{1, 10} = {-}{0.15em},
  vline{2, 3, 6, 7} = {-}{0.08em},
}
Strategy            & ICM500      & MLP-DP         & NCF            & DNN            & KuaiRand    & MLP-DP         & NCF            & DNN            \\
                    & ALL Feature & 0.1468         & 0.1388         & 0.1105         & ALL Feature & 0.0350         & 0.0343         & 0.0383         \\
PDQUBO (Random)     & 350         & 0.1445         & \uline{0.1460} & \uline{0.1110} & 179         & \uline{0.0347} & \uline{0.0347} & 0.0379         \\
remove\_jaccard~    &             & 0.2000         & 0.2048         & 0.1765         &             & 0.1250         & 0.1111         & 0.1842         \\
PDQUBO (Two\_Stage) &             & 0.1436         & 0.1401         & 0.1094         &             & 0.0338         & \uline{0.0347} & \uline{0.0390} \\
remove\_jaccard~    &             & 0.2712         & 0.2448         & 0.2448         &             & 0.1392         & 0.1392         & 0.1250         \\
PDQUBO (Full)       &             & 0.1465         & 0.1472         & 0.1166         &             & 0.0355         & 0.0355         & 0.0396         \\
PDQUBO (Random)     & 400         & 0.1480         & \uline{0.1450} & 0.1127         & 190         & \uline{0.0349} & \uline{0.0344} & \uline{0.0373} \\
remove\_jaccard~    &             & 0.2048         & 0.1628         & 0.1561         &             & 0.0625         & 0.0794         & 0.1724         \\
PDQUBO (Two\_Stage) &             & 0.1456         & 0.1423         & 0.1101         &             & 0.0347         & 0.0342         & 0.0367         \\
remove\_jaccard~    &             & 0.2270         & 0.1364         & 0.1765         &             & 0.1333         & 0.0794         & 0.0968         \\
PDQUBO (Full)       &             & \uline{0.1473} & 0.1491         & \uline{0.1107} &             & 0.0368         & 0.0352         & 0.0394         \\
PDQUBO (Random)     & 450         & 0.1438         & 0.1374         & \uline{0.1145} & 201         & 0.0346         & 0.0339         & 0.0382         \\
remove\_jaccard~    &             & 0.0989         & 0.0870         & 0.0417         &             & 0.0698         & 0.1220         & 0.0952         \\
PDQUBO (Two\_Stage) &             & \uline{0.1443} & \uline{0.1382} & 0.1139         &             & \uline{0.0361} & \uline{0.0348} & \uline{0.0387} \\
remove\_jaccard~    &             & 0.0989         & 0.0870         & 0.0638         &             & 0.0698         & 0.0952         & 0.1500         \\
PDQUBO (Full)       &             & 0.1476         & 0.1438         & 0.1232         &             & 0.0365         & 0.0353         & 0.0390         \\
PDQUBO (Random)     & 470         & 0.1453         & 0.1367         & 0.1103         & 212         & 0.0349         & \uline{0.0346} & 0.0374         \\
remove\_jaccard~    &             & 0.0714         & 0.0345         & 0.0909         &             & 0.0909         & 0.0909         & 0.0000         \\
PDQUBO (Two\_Stage) &             & \uline{0.1488} & \uline{0.1369} & \uline{0.1122} &             & \uline{0.0350} & 0.0342         & \uline{0.0383} \\
remove\_jaccard~    &             & 0.1765         & 0.0345         & 0.0714         &             & 0.0435         & 0.0435         & 0.0909         \\
PDQUBO (Full)       &             & 0.1499         & 0.1430         & 0.1177         &             & 0.0359         & 0.0351         & 0.0391         
\end{tblr}
}
\end{table}

\subsection{PDQUBO vs Traditional Feature Selection}
\label{sec:Ctfsm}

\begin{table*}[ht]
\centering
\caption{ Performance comparison between quantum algorithms and traditional feature selection algorithms in CTR tasks. Bold numbers indicate the best performance in each task, and underlined numbers indicate the second-best performance.}
\label{tab:table5}
\resizebox{0.85\linewidth}{!}{
\footnotesize
\begin{tblr}{
  width = \linewidth,
  colspec = {Q[90]Q[260]Q[69]Q[69]Q[69]Q[69]Q[69]Q[69]Q[69]Q[69]},
  cells = {c},
  cell{1}{1} = {r=2}{},
  cell{1}{2} = {r=2}{},
  cell{1}{3} = {c=2}{0.158\linewidth},
  cell{1}{5} = {c=2}{0.158\linewidth},
  cell{1}{7} = {c=2}{0.158\linewidth},
  cell{1}{9} = {c=2}{0.158\linewidth},
  cell{3}{1} = {r=15}{},
  cell{18}{1} = {r=15}{},
  vline{1} = {-}{0.15em},
  vline{2,3} = {-}{0.08em},
  vline{11} = {-}{0.15em},
  hline{3,18} = {-}{0.15em},
  hline{11,26} = {2-10}{0.08em},
  hline{1,33} = {-}{0.15em},
}
Base Models & Methods        & Avazu            &                  & Criteo           &                  & ICM\_150         &                  & ICM\_500         &                  \\
               &                & AUC              & Logloss          & AUC              & Logloss          & AUC              & Logloss          & AUC              & Logloss          \\
DeepFM         & no\_selection  & 0.74294          & 0.40105          & 0.76796          & 0.47816          & 0.84556          & 0.22260          & 0.84828          & 0.22176          \\
               & Lasso~\cite{tibshirani1996regression} & 0.74252          & 0.40094          & 0.76764          & 0.47869          & 0.84510          & 0.22765          & 0.85228          & 0.22090          \\
               & GBDT~\cite{friedman2001greedy} & 0.74284          & 0.40095          & 0.76806          & 0.47844          & 0.84542          & 0.22303          & 0.84839          & 0.22168          \\
               & AutoField~\cite{wang2022autofield} & 0.74318          & 0.40084          & 0.76831          & 0.47818          & 0.84510          & 0.22313          & 0.84806          & 0.22168          \\
               & LPFS~\cite{guo2022lpfs} & 0.74267          & 0.40099          & 0.76814          & 0.47824          & \textbf{0.84795} & \uline{0.22255}  & \textbf{0.85473} & \uline{0.22058}  \\
               & SFS~\cite{wang2023single} & 0.74296          & 0.40082          & 0.76852          & 0.47789          & 0.84618          & 0.22278          & 0.85212          & 0.22247          \\
               & Permutation~\cite{fisher2019all} & \textbf{0.74355} & \uline{0.40057}  & \uline{0.76855}  & 0.47816          & 0.84640          & 0.22263          & 0.84874          & 0.22160          \\
               & SHARK~\cite{zhang2023shark} & 0.74318          & 0.40060          & 0.76826          & \uline{0.47791}  & 0.84542          & 0.22306          & 0.85031          & 0.22131          \\
               & CQFS           & 0.74294          & 0.40105          & 0.76615          & 0.47946          & 0.84565          & 0.22285          & 0.85366          & 0.22056          \\
               & MIQUBO         & 0.74318          & 0.40084          & 0.76632          & 0.47954          & 0.84741          & 0.22272          & 0.85351          & 0.22049          \\
               & QUBO-Boosting  & 0.74266          & 0.40093          & 0.76770          & 0.47892          & 0.84201          & 0.22403          & 0.84657          & 0.22193          \\
               & CoQUBO         & 0.74254          & 0.40093          & 0.76703          & 0.47888          & 0.84405          & 0.22364          & 0.84694          & 0.22446          \\
               & PDQUBO\_Indiv	 & 0.74318 	        & 0.40060 	       & 0.76824 	      & 0.47799 	     & 0.84696 	        & 0.22276          & 0.85353 	      & 0.22062          \\
               & PDQUBO\_SA     & \uline{0.74333}  & \textbf{0.40048} & 0.76820          & 0.47824          & 0.84695          & 0.22278          & 0.85317          & 0.22063          \\
               & PDQUBO\_hybrid & \uline{0.74333}  & \textbf{0.40048} & \textbf{0.76857} & \textbf{0.47770} & \uline{0.84788}  & \textbf{0.22240} & \uline{0.85456}  & \textbf{0.22039} \\
FiBiNET        & no\_selection  & 0.74169          & 0.40148          & 0.77003          & 0.47697          & 0.83319          & 0.22802          & 0.83731          & 0.22558          \\
               & Lasso~\cite{tibshirani1996regression}          & 0.74114          & 0.40183          & 0.76972          & 0.47730          & 0.84241          & 0.22588          & 0.84503          & 0.22392          \\
               & GBDT~\cite{friedman2001greedy}           & 0.74161          & 0.40157          & 0.77134          & 0.47605          & 0.84029          & 0.22631          & 0.84659          & 0.22334          \\
               & AutoField~\cite{wang2022autofield}      & 0.74161          & 0.40157          & 0.77141          & 0.47591          & 0.84150          & 0.22614          & 0.84584          & 0.22321          \\
               & LPFS~\cite{guo2022lpfs}           & 0.74159          & 0.40158          & 0.77146          & 0.47584          & \uline{0.85546}  & \textbf{0.22172} & \textbf{0.84883} & \textbf{0.22268} \\
               & SFS~\cite{wang2023single}            & 0.74161          & 0.40157          & 0.77099          & 0.47621          & 0.85295          & 0.22271          & 0.84456          & 0.22336          \\
               & Permutation~\cite{fisher2019all}    & \uline{0.74209}  & \uline{0.40138}  & \uline{0.77186}  & 0.47598          & 0.84050          & 0.22636          & 0.84220          & 0.22454          \\
               & SHARK~\cite{zhang2023shark}          & 0.74193          & 0.40146          & 0.77099          & 0.47621          & 0.83937          & 0.22670          & 0.84542          & 0.22369          \\
               & CQFS           & 0.74185          & 0.40154          & 0.77055          & 0.47646          & 0.83569          & 0.22771          & 0.22056          & 0.22471          \\
               & MIQUBO         & 0.74193          & 0.40146          & 0.77021          & 0.47698          & 0.84825          & 0.22414          & 0.84364          & 0.22444          \\
               & QUBO-Boosting  & 0.74191          & 0.40166          & 0.76906          & 0.47788          & 0.83569          & 0.22771          & 0.84306          & 0.22445          \\
               & CoQUBO         & 0.74169          & 0.40148          & 0.77060          & 0.47669          & 0.83886          & 0.22675          & 0.84028          & 0.22647          \\
               & PDQUBO\_Indiv	 & 0.74217 	        & 0.40131 	       & 0.77060 	      & 0.47669 	     & 0.85329 	        & 0.22255 	       & 0.84744 	      & 0.22317          \\
               & PDQUBO\_SA     & \textbf{0.74217} & \textbf{0.40131} & 0.77157          & \uline{0.47583}  & 0.85511          & 0.22206          & 0.84771          & 0.22318          \\
               & PDQUBO\_hybrid & \textbf{0.74217} & \textbf{0.40131} & \textbf{0.77197} & \textbf{0.47561} & \textbf{0.85567} & \uline{0.22182}  & \uline{0.84864}  & \uline{0.22269}  
\end{tblr}
}
\end{table*}

Here, we examine the performance differences between Quantum-based QUBO algorithms (including PDQUBO) \footnote{ In CTR datasets, where no explicit user–item interaction matrix is available, we adopt a representation-based adaptation of CQFS. Specifically, we compute item-level similarity in the learned embedding space of the CTR model and treat it as a latent collaborative signal, while content similarity is derived from the raw feature space. The alignment between these two similarity structures is incorporated into the QUBO coefficient construction, following the core principle of CQFS.} 
and traditional feature selection methods implemented on classical computers. Although current quantum hardware still faces significant limitations in solving real-world problems, making such comparisons less than completely fair, it is still worth exploring. Considering that feature selection tasks in recommender systems are primarily applied to CTR scenarios, we incorporated two widely used CTR datasets, Avazu and Criteo, and applied a 1:5 negative sampling strategy to the ICM dataset to better align it with CTR tasks. Additionally, we replaced the base models with DeepFM~\cite{guo2017deepfm} and FiBiNET~\cite{huang2019fibinet}, both implemented using the DeepCTR framework, and changed the evaluation metric from nDCG to AUC (Area under the curve) to better suit the CTR context. The dataset partitioning follows the approach outlined in Sec.~\ref{sec:dataset}. For traditional feature selection methods, we employed the benchmark proposed in ERASE~\cite{jia2024erase} with default parameters to generate feature rankings. For QUBO algorithms on quantum architectures, to reduce the randomness inherent in solving quadratic problems, we repeated the selection process five times and adopted the feature selection results corresponding to the smallest $Y$ value. Subsequently, the selected features were used to retrain the base models, instead of reusing the pre-trained model parameters as in previous experiments. To be fair, all hyperparameters of the retrained base models were kept consistent, including a fixed batch size of 1024 and an L2 regularization coefficient of 1e-5.

In Table~\ref{tab:table5}, it is observed that PDQUBO\_hybrid consistently achieves the best or second-best performance across all scenarios 
(Note, according to the benchmark in feature selection~\cite{jia2024erase,zhang2023shark}, even relatively small performance gains are considered meaningful). This demonstrates that, empowered by the precise and efficient optimization capabilities of quantum annealers, PDQUBO\_hybrid not only matches or surpasses traditional feature selection methods in performance but also exhibits greater stability across diverse datasets and models. In particular, PDQUBO significantly outperforms all QUBO-based baseline methods, confirming the effectiveness of the proposed performance-driven construction of the Q matrix. Compared to sensitivity-based methods such as Permutation, PDQUBO achieves substantial performance gains, especially on feature-rich datasets like ICM, where Permutation struggles. This improvement can be attributed to PDQUBO’s ability to capture pairwise feature interactions during the selection process, as encoded in the quadratic terms of the QUBO matrix. Moreover, PDQUBO\_hybrid consistently outperforms or matches PDQUBO\_SA, highlighting the value of quantum annealers in more effectively exploring high-quality feature subsets. Notably, although the counterfactual instances ${E}_i$ and ${E}_{ij}$ in Equation~\ref{eq:Equationd} are derived during inference using pre-trained base models, the final recommendation models are retrained with the selected features, ensuring generalizability of the selected subsets. The variant PDQUBO\_Indiv, which only uses diagonal elements in the Q matrix and ignores pairwise feature effects, results in degraded performance. This further underscores the importance of incorporating interaction terms when constructing the QUBO formulation for feature selection. Finally, prior work such as ERASE~\cite{jia2024erase} has shown that no single feature selection method consistently outperforms all others across every scenario. In contrast, PDQUBO\_hybrid demonstrates robust and competitive performance compared to both quantum and classical baselines, further validating its practical utility in recommender systems.

\section{CONCLUSION}\label{sec:conclusion}

{This paper proposes PDQUBO, a performance-driven QUBO formulation tailored for feature selection in recommender systems on quantum annealers. Experimental results demonstrate that PDQUBO outperforms state-of-the-art baselines in terms of recommendation accuracy across various models and datasets. Beyond effectiveness, we further investigate the robustness of quantum annealing for recommendation problems by analyzing the stability of solution quality under repeated runs, and explore the necessity of quadratic interaction terms within QUBO formulations.

Looking forward, future work could extend PDQUBO to support higher-order feature interactions beyond pairwise dependencies, enabling the modeling of more complex feature structures through quantum optimization. Moreover, although our current implementation is based on quantum annealers, the underlying formulation is compatible with general-purpose gate-based quantum computers, opening up opportunities for exploring variational quantum algorithms or hybrid quantum-classical solvers in large-scale recommender systems.

\begin{acks}

This research is supported by the ARC Discovery Project (DP210100743) and RACE (RMIT Advanced Cloud Ecosystem). We would like to express our sincere gratitude to RACE for providing access to quantum computing hardware and computational infrastructure, which was essential for conducting our experiments and advancing this research.

\end{acks}
\

\bibliographystyle{ACM-Reference-Format}
\bibliography{sample-base}

\appendix

\end{document}